\documentclass[amsmath,amssymb,aps,showpacs,floatfix,pre,twocolumn,longbibliography]{revtex4-1}
\usepackage{graphicx}
\usepackage{bm}
\usepackage{hyperref}
\usepackage{color,array}

\begin{document}

\title{Recurrence dynamics of particulate transport with reversible blockage:\\  from a single channel to a bundle of coupled channels}
\author{Chlo\'e Barr\'e, Gregory Page, Julian Talbot and Pascal Viot}
\affiliation{
Laboratoire de Physique Th\'eorique de la Mati\`ere Condens\'ee, SU, CNRS  UMR 7600,
 4, place Jussieu, 75005 Paris Cedex 05, France
}
\begin{abstract}
We model a particulate flow of constant velocity through confined geometries, ranging from a single
channel to a bundle of $N_c$ identical coupled channels, under conditions of reversible blockage.
Quantities of interest include the exiting particle flux (or throughput) and the probability that the
bundle is open. For a constant entering flux, the bundle evolves through a transient regime to a
steady state. We present analytic solutions for the stationary properties of a
single channel with capacity $N\le 3$ and for a bundle of channels each of capacity $N = 1$. For
larger values of $N$ and $N_c$, the system's steady state behavior is explored by numerical
simulation. Depending on the deblocking time, the exiting flux either increases monotonically with intensity or
displays a maximum at a finite intensity. For large $N$ we observe an abrupt change from a state with
few blockages to one in which the bundle is permanently blocked and the exiting flux is due entirely
to the release of blocked particles. We also compare the relative efficiency of coupled and uncoupled bundles. 
For $N=1$ the coupled system is always more efficient, but for $N>1$ the behavior is more complex.
\end{abstract}
\date{\today}
\maketitle

\section{Introduction}

Many physical and biological processes feature particle flow in confined geometries. Examples include vehicular and pedestrian traffic \cite{Helbing2003,Appert-Rolland2010a}, filtration of particle suspensions,  and the flow of macromolecules or ions through micro- or nanochannels  \cite{Finkelstein1981,Kelkar20072011,daiguji2004}. In biology, an important example is flux regulation mediated by pore-forming membrane proteins. The transport of ions and water through these channels is primarily a single-file process, i.e. cations and water molecules cannot pass each other within the channel \cite{daiguji2004}. Moreover, these flux regulatory channels can be clogged by toxins or medicines, with significant consequences. Recent studies of tracer diffusion of {\it biased} (or active) particles  in a crowded, narrow channel revealed a non-trivial relationship between the exerted force and the mean velocity, as well asymmetric density profiles of the environment
\cite{0953-8984-30-44-443001,Bertrand2018,Krapivsky2014,Benichou2018}.

Some information systems, such as telecommunication \cite{Senderovich2015278} and computing networks \cite{6745901,ezaki2015taming}, as well as trunked mobile radio systems and air traffic \cite{janssen2008, barcelo1996,6820567} are also amenable to the channel description. 

A blockage may be caused by either `extrinsic' or `intrinsic' mechanisms. The former refers to the situation where the number of particles present somehow exceeds the channel carrying capacity, and will be the focus of this article. The latter mechanism arises from collective effects such as encountered in filtration processes. In this case, while isolated particles can pass through a mesh hole, clogging occurs when two or more particles arrive in near concurrence, causing one to impede the other. This effect, due to the delicate interplay between the spatio-temporal closeness of the particles and the confining geometry, could be seen as setting the capacity of a channel to greater than one, which establishes a connection between both types of blocking mechanisms. A model based on this phenomenology successfully accounted for experimental data \cite{Roussel2007,Redner2000a}. 
Various approaches, including the Totally Asymmetric Simple Exclusion Process (TASEP)\cite{Mallick2011,Mallick2015}  have been applied to model these phenomena. Gabrielli \textit{et al.} introduced a class of stochastic models in which blockages occur when the carrying capacity of a channel is exceeded \cite{Gabrielli2013,talbot2015}. For these stochastic models, the particle velocity within the channels is identical, and the mean particle density is low enough to prevent exclusion effects. The blockage is triggered when the  number of particles within the channel at a given time exceeds the channel capacity.  The original model considered one channel with capacity $N=2$, i.e., two particles must be simultaneously present in the channel to block the system.  Particles enter at random times according to a Poisson process of intensity $\lambda$ and exit, if no blockage occurs,
after a fixed transit time $\tau$. Subsequently, several generalizations were studied, including a higher blocking threshold ($N>2$) \cite{barre2015a}, an inhomogeneous entering flux \cite{barre2015}, and multiple channels \cite{PhysRevE.92.052141,barre2016diminishing}. 
When the blockage is reversible, the system is reactivated after a constant waiting time, $\tau_b$. This mechanism gives rise to 
a transient regime leading to a steady state \cite{Barre2013}. In this article we associate the last two generalizations.

Queuing theory \cite{Adan2002,Medhi:1991:SMQ:113159} provides an alternative description of reversible blocking phenomena. This framework is typically used to describe customers arriving at a server, according to one given distribution, and receiving service according to another distribution. If all the elementary steps are Markov processes, the time evolution of the state probabilities can be described by systems of linear differential equations.  In accordance with this approach, we recently introduced Markovian models of blockage \cite{1742-5468-2018-6-063213,0953-8984-30-30-304004} for which exact solutions can be obtained.

This paper is organized as follows:
We first consider the single channel model in Sec. \ref{sec:onechannel}, for which exact solutions are available for $N=1,2,3$, and simulation results for greater values of $N$. We also compare our results with the previously introduced Markovian models. In Sec. \ref{sec:multimodel} a channel bundle model consisting of $N_c$ identical channels each of capacity $N$ and sharing an incoming flux is presented.  Exact results are obtained for $N = 1$, as well as numerical simulation results for larger values of $N$.  These allow us to observe and explain some results in the limit of large $N$ or $N_c$.   In Sec.  \ref{sec:compare}, we compare the efficiency of different configurations of channels in conveying a particulate flux of given intensity. 
Finally, in Sec. \ref{sec:discussion} we summarize our results.

\section{Single channel Model}
\label{sec:onechannel}

Particles with identical constant velocities, are injected into a channel of length $L$, according to a Poisson distribution of mean intensity $\lambda$. Given no blockage occurs, the particle transit time is $\tau=L/v$. An instantaneous blockage occurs if $N$ particles are simultaneously present, and lasts for time $\tau_b > \tau$. In the limit $\tau_b\rightarrow \infty$, there is no steady state and the exiting flux falls to zero \cite{barre2015a}. Here, we focus instead on reversible blockages, during which, the $N$ particles are retained, and no more may enter the channel.  After the deblocking time, the channel instantaneously releases all $N$ particles, resetting to the empty state, thereby allowing new ones to enter. The dynamics is therefore a recurring cycle of alternating open and closed states, that ultimately leads to a stationary state.

In an average steady state recurrence cycle, the channel is open for an average time $\langle t\rangle$ and blocked for a fixed time $\tau_b$.  The stationary probability that the system is open is therefore
\begin{equation}
p_{o}(\lambda)=\frac{\langle t \rangle }{\langle t \rangle+\tau_b},
\label{eq:Po}
\end{equation}
where the denominator represents the total mean time of a recurrence. The stationary output flux is then given by the ratio of the mean number of particles released during one cycle to the cycle period,
\begin{equation}
j(\lambda)=\frac{\langle m \rangle +N}{\langle t \rangle+\tau_b}.
\label{eq:j}
\end{equation}
$\langle m\rangle$ is the mean number of output particles between two successive blockages. By equating the number of entering particles in one period to those exiting we obtain the following `number balance'
\begin{equation} \label{eq:mt}
\langle m \rangle +N=\lambda \langle t\rangle.
\end{equation}
Finally from the above three equations we deduce that
\begin{equation} 
j(\lambda)=\lambda p_{o}(\lambda).
\label{eq:jpo}
\end{equation}
The latter relation does not depend on the existence of a cycle, as it is the result of number conservation. The output flux is equal to the entering one minus the part that is rejected when the channel is in the closed state. 

By taking the limit $\lambda\tau\ll 1$, the mean blockage time and the mean number of exiting particles between blockages behave asymptotically  as $\langle t\rangle\gg \tau_b$ and $\langle m\rangle\gg N$, respectively. Therefore, for a given $\tau_b$, the probability that the channel is open is close to unity and the flux $j(\lambda)\simeq \lambda$. Blockages rarely occur at low $\lambda$. In this limit, the mean blockage time can be estimated by noting that a blockage occurs when a batch of particles enters in a finite duration $\tau$, 
leading to $\langle t\rangle= \tau \frac{(N-1)!}{(\lambda\tau)^N}$ \cite{barre2015a}. Expanding  $p_o(\lambda)$ to first order gives
\begin{equation}
p_{o}(\lambda)\simeq\frac{1}{1+\frac{(\lambda\tau)^N\tau_b}{(N-1!)\tau}} .
\label{eq:Pol}
\end{equation}
When $\lambda\tau\gg 1$, blockages are very frequent, and both the mean number of exiting particles between blockages, $\langle m\rangle$, and mean time between blockages, $\langle t\rangle$, approach zero. The resulting flux consists entirely of successive releases of the blocked particles, $j(\lambda)=N/\tau_b$ and $p_o(\lambda)\simeq 0$. In this limit, $\langle t \rangle$ corresponds to the time necessary for $N$ particles to enter an empty channel, $N/\lambda$.
The open probability and the flux in this high intensity limit are therefore
\begin{equation}
p_{o}(\lambda)\simeq\frac{N}{N+\lambda\tau_b},
\label{eq:Poh}
\end{equation}
and
\begin{equation}
j(\lambda)\simeq \frac{N\lambda}{N+\lambda\tau_b}.
\label{eq:j1}
\end{equation}

\subsection{Solvable models: $N\leq 3$}
\label{sec:RN1}
For small capacities, $N \in [1,2,3] $, the time evolution of the process can be expressed by analytically tractable differential or integro-differential equations \cite{Barre2013}. For larger values of $N$, the time evolution cannot be solved by any known means.

We first consider $N=1$, which corresponds to a stochastic switch. The transit time $\tau$ is an irrelevant variable because no particle can exit the channel without having already blocked it. For $N>1$, it is possible for particles to pass through the channel without causing a blockage. Let $p_o(t)$ denote the probability that the channel is open  at time $t$. Its time evolution obeys
\begin{equation}\label{eq:pON1}
\frac{dp_o(t)}{dt}=-\lambda p_o(t)+\lambda p_o(t-\tau_b).
\end{equation}
The loss term corresponds to the entrance of a particle at time $t$, while the channel is open, causing the channel to block. The gain term corresponds to the exit of a particle that became blocked at time $t-\tau_b$, with the subsequent reopening of the channel at time $t$.

The mean output flux at time $t$ is given by
\begin{equation}\label{eq:jtN1}
j(t)=\lambda p_o(t-\tau_b),
\end{equation}
which corresponds to the release of a blocked particle that entered at $t-\tau_b$. Applying the time Laplace transform, $\tilde{f}(u)=\int_0^\infty dt e^{-ut}f(t)$, to Eqs.(\ref{eq:pON1}) and (\ref{eq:jtN1}) gives,
\begin{equation}\label{eq:Pou}
\tilde{p}_o(u)=\frac{1}{\lambda+u-\lambda e^{-u\tau_b}},
\end{equation}
and 
\begin{equation}\label{eq:jou}
\tilde{j}(u)= \frac{\lambda e^{-u\tau_b} }{\lambda+u-\lambda e^{-u\tau_b}}.
\end{equation}

Expanding the denominator of Eq.(\ref{eq:Pou}) in terms of $\lambda e^{-u\tau_b}/(\lambda+u)$, allows one to easily invert the Laplace transform, term by term, giving,

\begin{equation}
 p_o(t)=\sum_{n=0}^{\infty}\frac{[\lambda(t-n\tau_b)]^n}{n!}e^{-\lambda(t-n\tau_b)}\theta(t-n\tau_b),
\end{equation}
where $\theta(t)$ is the Heaviside function. The stationary open probability, $p_o(\lambda)$, and flux, $j(\lambda)$, can be obtained from Eqs.(\ref{eq:Pou}) and (\ref{eq:jou}) by using $\tilde{f}(u)\simeq \frac{f(\lambda)}{u}$,
\begin{equation}\label{eq:Polo}
p_ {o}(\lambda)=\frac{1}{1+\lambda\tau_b},
\end{equation}
and 
\begin{equation}\label{eq:jolo}
j(\lambda)=\frac{\lambda}{1+\lambda\tau_b}.
\end{equation}
These results can be easily inferred from Eqs. (\ref{eq:Po}) and (\ref{eq:j}) by setting  $\langle m\rangle=0$ and $\langle t\rangle=1/\lambda$.
The exiting particle flux is controlled by the incoming flux $\lambda$ and the time of blockage $\tau_b$. 

We now consider the $N=2$ model, i.e. blockage occurs when two particles are simultaneously in the channel, for which exact results have already been obtained \cite{Barre2013}. Here we propose an alternative, simpler derivation using the state probabilities of the channel. Let $p_0(t)$, $p_1(t)$ denote the probability that an open channel contains zero or one particle respectively, and $p_2(t)$ be the probability that it contains two particles and is therefore blocked. The time evolution of the process is given by 
\begin{align}
\frac{d p_{0}}{dt}&=-\lambda p_{0}(t) +\lambda e^{-\lambda \tau} p_{0} (t-\tau) +\lambda p_{1}(t-\tau_b), \label{eq:n2a}\\
\frac{d p_{1}}{dt}&= -\lambda e^{-\lambda \tau} p_{0} (t-\tau) -\lambda p_{1}(t)+\lambda p_{0}(t) \label{eq:n2b},\\
\frac{d p_{2}}{dt}&=-\lambda p_{1}(t-\tau_b) +\lambda p_{1}(t),\label{eq:n2c}
\end{align}
with the following initial conditions:
\begin{equation}
p_0(0) = 1, \; p_1(0) = p_2(0) = 0 . \label{eq:n2d}
\end{equation}

In Eq.(\ref{eq:n2a}), the loss term corresponds to the entrance of a particle in the empty channel at time $t$.  The two gain terms $\lambda e^{-\lambda \tau} p_{0} (t-\tau)$ and $\lambda p_{1}(t-\tau_b)$ correspond to a particle exiting the channel at time $t$  and a channel release (with a blockage occurring at time $t-\tau_b$), respectively. In Eq.(\ref{eq:n2b}), the two loss terms describe either a particle exiting the occupied channel at time $t$ or a particle entering the occupied channel. The gain term corresponds to a particle entering a free channel. In Eq.(\ref{eq:n2c}), the loss term corresponds to a channel release and the gain term to a particle entering a channel with one particle already inside. Summing the three equations verifies that the total probability is conserved: $p_{0}(t)+p_{1}(t)+p_{2}(t)=1$.

Taking the Laplace transform of Eqs.(\ref{eq:n2a}-\ref{eq:n2c}) gives
\begin{align}
\left[u+\lambda(1-e^{-\tau(u+\lambda) } )\right]\tilde{p}_{0}(u)-\lambda e^{-u\tau_b} \tilde{p}_{1}(u)&=1,\\
 - \lambda (1-e^{-\tau(u+\lambda)}) \tilde{p}_{0}(u)+(\lambda+u) \tilde{p}_{1}(u) &=0,\\
-\lambda(1-e^{-u\tau_b}) \tilde{p}_{1}(u)+u\tilde{p}_{2}(u)&=0.
\end{align}
\begin{figure}[!t]
	\begin{center}
		\includegraphics[width=8cm]{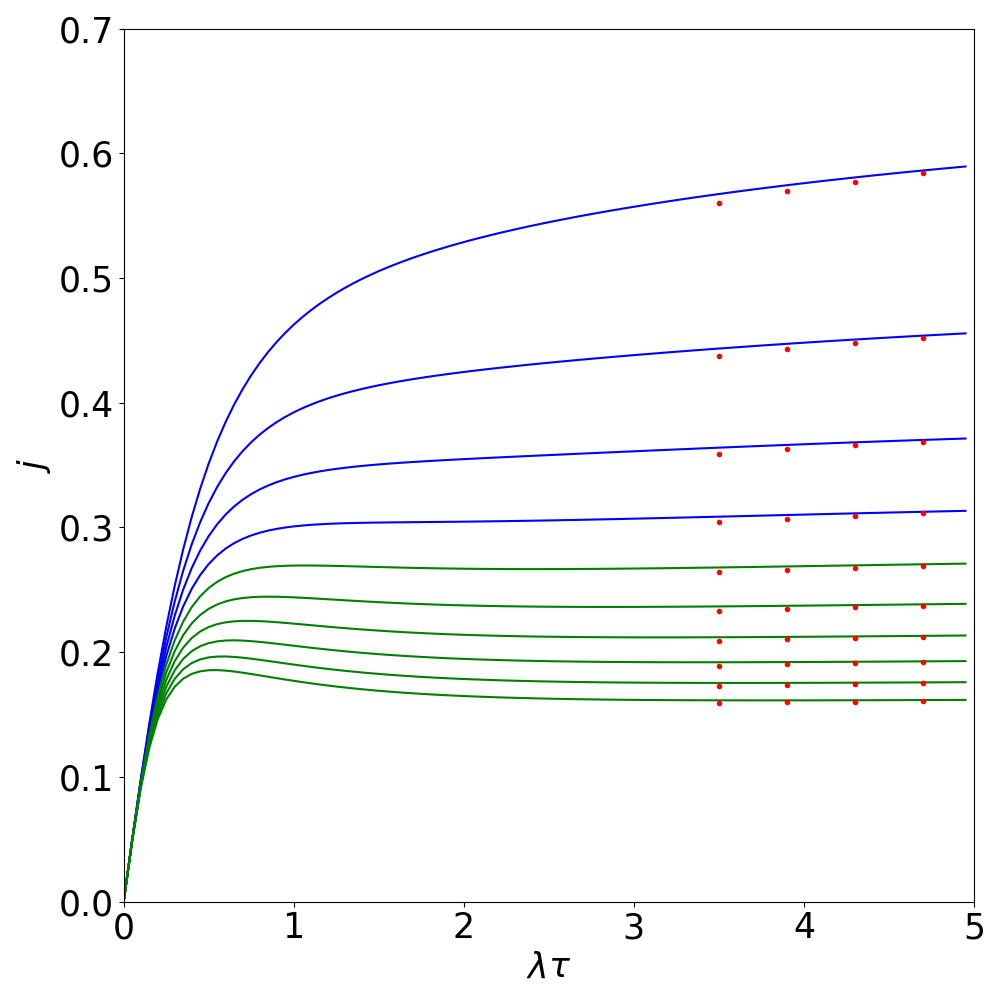}
	\end{center}
	\caption{  Exiting flux, $j(\lambda)$, versus $\lambda$ for $N=2$ and $\tau_b=3,4.\cdots,12$ from top to bottom.  The blue curves show cases where the flux increases monotonically with $\lambda$. The green curves show cases where the flux displays a maximum at a finite value of $\lambda$. The dotted lines correspond to the asymptotic values of the flux, Eq. (\ref{eq:j1} ). }
	\label{fig:NC1exact}
\end{figure}

These simultaneous equations may be solved to give

\begin{align} \label{eq:Pou1}
\tilde{p}_{0}(u)&=\frac{u+\lambda}{\Delta},   \\ \label{eq:Pou2}
\tilde{p}_{1}(u)&=\frac{\lambda}{\Delta}  (1-e^{-\tau (u+\lambda)}), \\
\tilde{p}_{2}(u)&=\frac{\lambda ^2}{u\Delta}(1-e^{-u\tau_b})(1-e^{-\tau (u+\lambda)}),
\end{align}
where
\begin{equation}
\Delta= (u+\lambda)^2-\lambda(\lambda +u)e^{-\tau (u+\lambda)}-\lambda^2 e^{-u\tau_b}(1-e^{-\tau (u+\lambda)}).
\end{equation}

The mean exiting flux $j(t)$ is the sum of two contributions: the exit of a particle from an open channel and the release of of two particles from a closed channel. $j(t)$ is therefore given by
\begin{equation}
j(t)=\lambda e^{-\lambda \tau} p_0(t-\tau)+2\lambda p_1(t-\tau_b).
\end{equation}
By using Eqs.(\ref{eq:Pou1}),(\ref{eq:Pou2}), the Laplace transform of the output flux $\tilde{j}(u)$ is 
\begin{equation}\label{eq:jdeu}
\tilde{j}(u)=\frac{\lambda}{\Delta}[(u+\lambda)e^{-(u+\lambda)\tau} +2\lambda e^{-u\tau_b}(1-e^{-\tau (u+\lambda)})].
\end{equation}
As expected, we recover the results of Ref.\cite{Barre2013} and the time-dependent mean flux can be obtained by a Laplace inversion of Eq.(\ref{eq:jdeu}).

We here focus on the key quantities, namely the stationary probability $p_o(\lambda)$ that the system is open and the mean flux $j(\lambda)$. $p_o(\lambda)$ is the sum of the two stationary probabilities $p_0$ and $p_1$, each obtained by evaluating
$\lim_{u\rightarrow 0}u \tilde{p}_i(u)$ with $i=0,1$:

\begin{equation}
 p_o(\lambda)= \frac{2 -e^{-\lambda  \tau }}{2+\lambda \tau_b -(1+\lambda \tau_b)e^{-\lambda  \tau}},
\end{equation}
and  
\begin{equation}\label{eq:jN2}
j(\lambda)=\lambda \frac{2 -e^{-\lambda  \tau }}{2+\lambda \tau_b -(1+\lambda \tau_b)e^{-\lambda  \tau}}.
\end{equation}

Figure \ref{fig:NC1exact} displays $j(\lambda)$ versus $\lambda\tau$ for different integer values of $\tau_b$ from $3$ to $12$. The dashed lines correspond to the asymptotic values of the exiting flux $\lim_{\lambda->\infty}j(\lambda)=\frac{2}{\tau_b}$. One first observes that the stationary flux reaches the asymptotic values more rapidly as $\tau_b$ increases. Moreover, $j(\lambda)$ displays a maximum when $\tau_b$ is larger than $7$. It is possible to obtain the exact value of $\tau_b$ for which the flux $j(\lambda)$ displays a maximum at a finite value of $\lambda$\cite{1742-5468-2018-6-063213} by solving $\frac{\partial j(\lambda)}{\partial \lambda}=0$. A real solution for $\lambda$ exists if $\tau_b>6.2$. Note that for $N=1$ the flux is always a monotonically increasing function of $\lambda$. 

For $N=3$, the complete kinetic description of the model is cumbersome so we restrict our attention to the stationary quantities for which analytical expressions have been obtained \cite{barre2015a}. In particular, the mean time to blockage starting from an empty channel is given by 
	\begin{equation}
	\lambda \langle t\rangle= \frac{2e^{\nu}\sinh(g\nu)+g e^{\lambda\tau}}{-g-2\sinh(g\nu)
		e^{-\nu}+e^{\nu}\left(\sinh(g\nu)+g\cosh(g\nu)\right)}+1\label{eq:temps}
	\end{equation}
for $\lambda\tau>2\ln(2)$ and 
	\begin{equation}
	\lambda \langle t\rangle= \frac{2e^{\nu}\sin(g\nu)+g e^{\lambda\tau}}{-g-2\sin(g\nu)
		e^{-\nu}+e^{\nu}\left(\sin(g\nu)+g\cos(g\nu)\right)}+1\label{eq:temps2}
	\end{equation}
for  $\lambda\tau<2\ln(2)$, where $g=\sqrt{|1-4e^{-\lambda\tau}|}$ and $\nu=\frac{\lambda \tau}{2}$ (note that these correct the expressions given in \cite{barre2015a}).
\begin{figure}[th]
	\begin{center}
		\includegraphics[width=8cm]{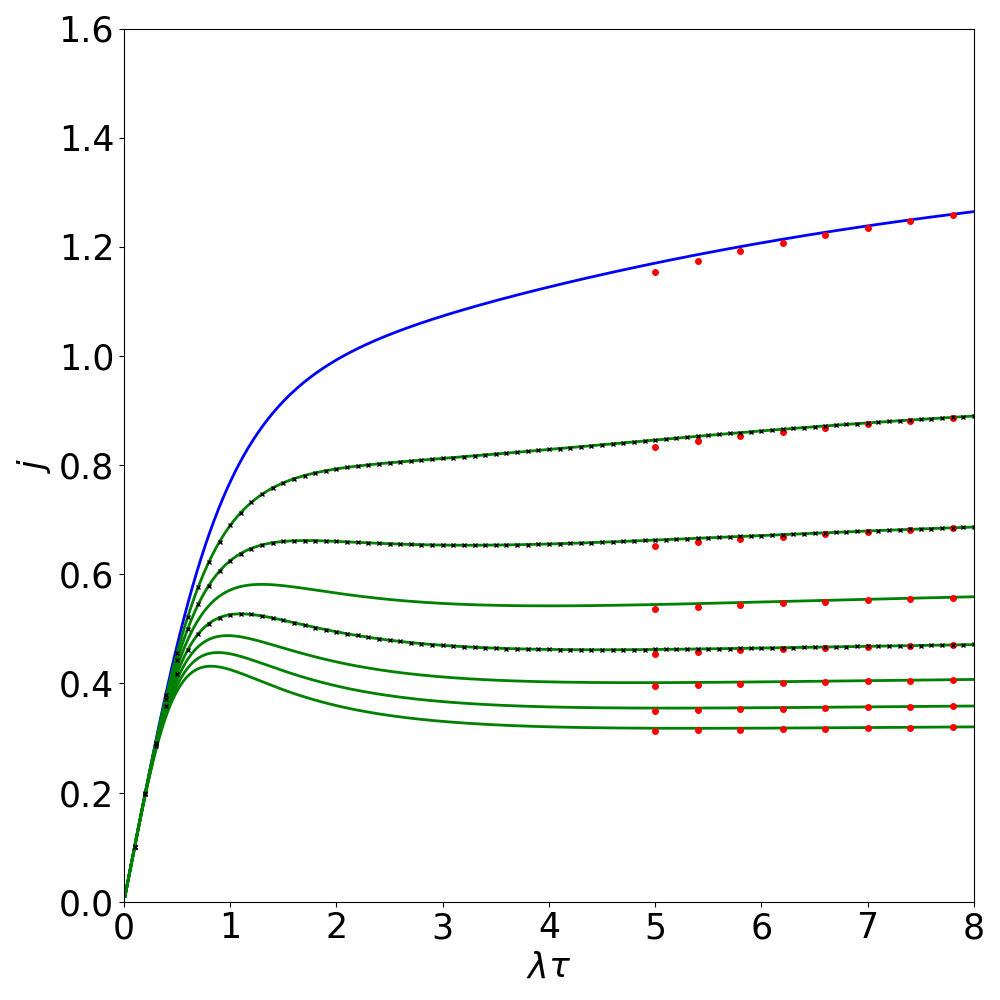}
	\end{center}
	\caption{  Exiting flux, $j(\lambda)$ versus $\lambda$ for $N=3$ and $\tau_b=2,3,\cdots,9$ from top to bottom. The blue curves indicate cases where the flux increases monotonically with $\lambda$. The green curves indicate the cases where the fluxes display a maximum at a finite value of $\lambda$. The black crosses show simulation results for $\tau_b=3,4,6$, which match perfectly with the exact results. The dotted lines indicate the asymptotic values of the flux, Eq. (\ref{eq:j1}). 
}
	\label{fig:NC1exact2}
\end{figure}
The two stationary quantities  $p_{o}(\lambda)$ and  $j(\lambda)$ are obtained by inserting this result in Eqs.(\ref{eq:Po}) and(\ref{eq:j}).
Fig. \ref{fig:NC1exact2} displays $j(\lambda)$ as a function of $\lambda\tau$ for different values of $\tau_b$. There are several differences compared with the $N=2$ model. First, a maximum exiting flux occurs if the blockage time $\tau_b>3.6$, which is significantly smaller for $N=2$ ($\tau_b>6.2$). Second, the asymptotic values are reached at a lower value of $\lambda$, and finally, the intensity $\lambda$ at which $j(\lambda)$ is maximum is also shifted towards larger intensity.

\subsection{Simulation results: $N> 3$}
\begin{figure}[th]
	\begin{center}
		\includegraphics[width=8cm]{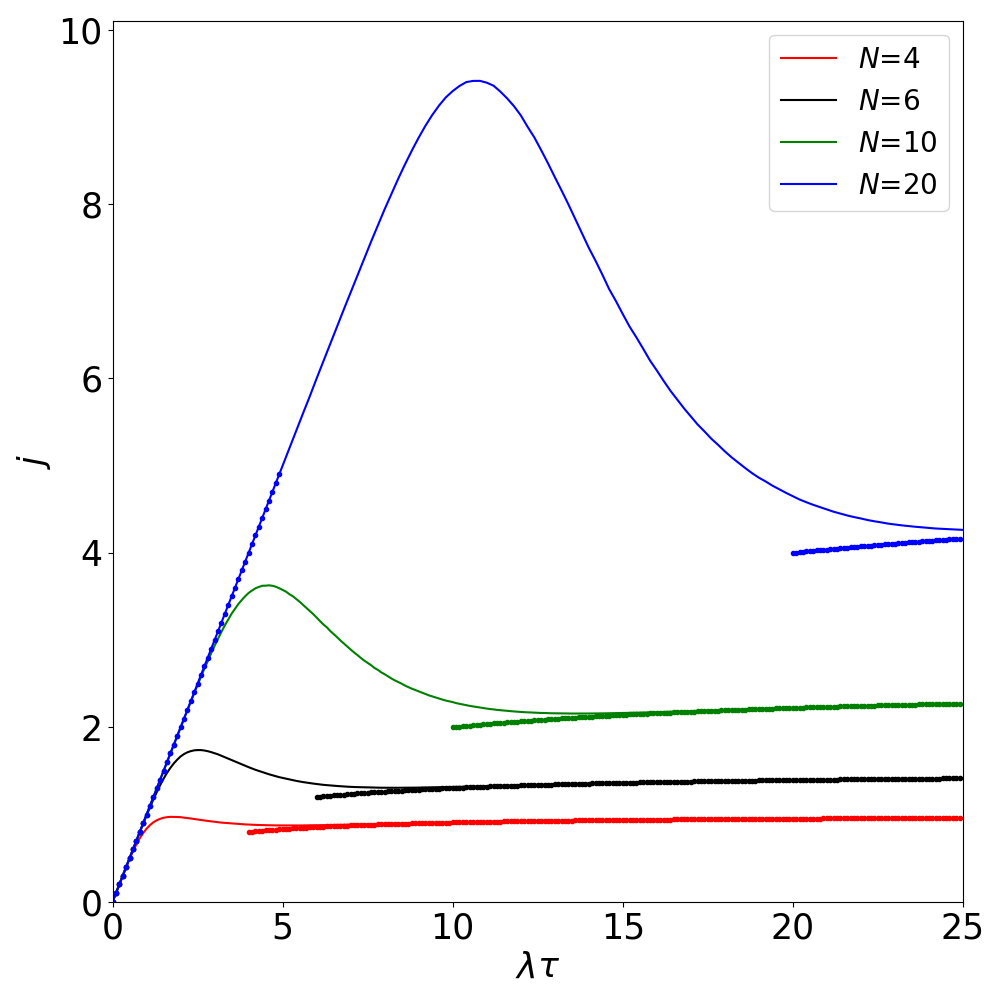}
	\end{center}
	\caption{Exiting particle flux, $j(\lambda)$,  versus $\lambda$ for $N=4,6,10,20$ and $\tau_b=4$. Dotted curves correspond to the asymptotic values at low and high intensity.}
	\label{fig:NC1}
\end{figure}

\begin{figure}[th]
	\begin{center}
		\includegraphics[width=8cm]{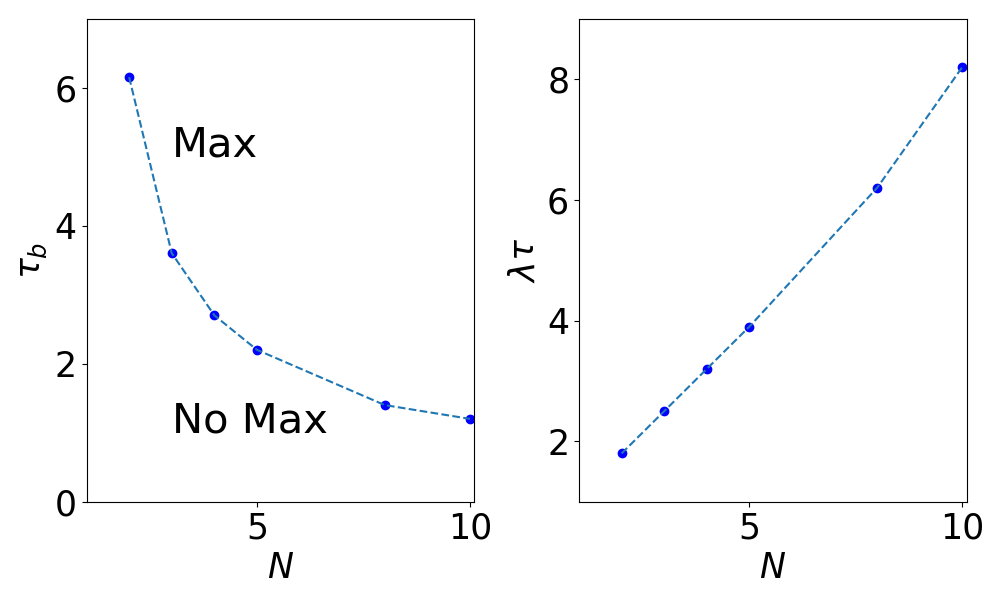}
	\end{center}
	\caption{ (left) Critical values of $\tau_b$ as a function of $N$ above which the steady state flux of a single channel displays a maximum at finite $\lambda$. (right) The corresponding values of $\lambda\tau$. }
	\label{fig:taubNC1}
\end{figure}

\begin{figure}[t!]
	\begin{center}
		\includegraphics[width=7cm]{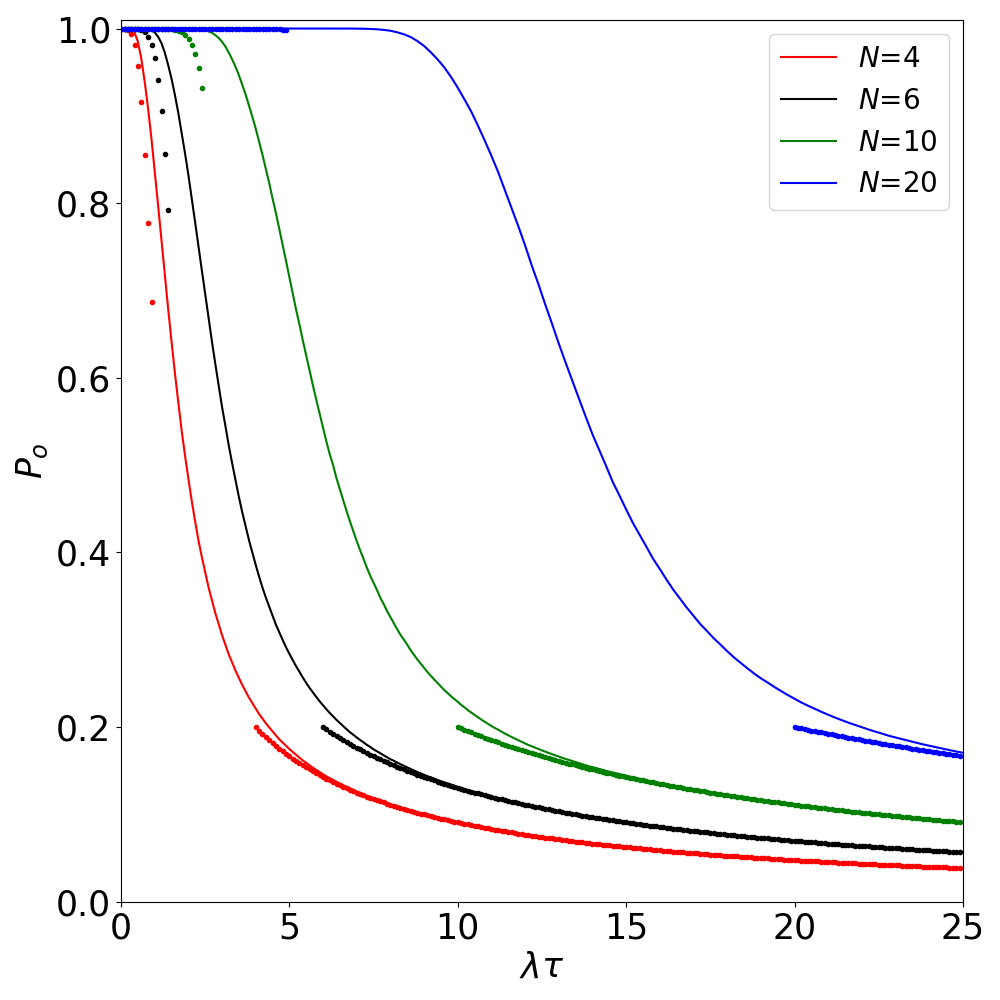}
	\end{center}
	\caption{Open probability $p_o$ versus $\lambda\tau$ for $N=4,6,10$ and $\tau_b=4$. Dotted curves correspond to the asymptotic values at low and high intensity.}
	\label{fig:NC1proba}
\end{figure}
As a result of strong time correlations between the transiting particles, it is not possible to obtain analytic solutions for $N>3$. 
We therefore used numerical simulations to investigate these cases.
In order to benchmark our code, we compared the simulation results for the stationary flux for $N=3$ with the exact expressions for three different values of $\tau_b$. In Fig. \ref{fig:NC1exact2} we observe perfect agreement between the analytical expressions and the simulation results.

Figure \ref{fig:NC1} displays the stationary exiting flux as a function of the intensity for different values of $N$ and $\tau_b=4$. At low intensity, the flux increases linearly, and at high intensity the asymptotic behavior of the simulation results is well-desribed by Eq. (\ref{eq:j1}). The behavior in the intermediate region is due to complex dynamics that alternates between blockages and sequences of uninterrupted transport. For $N > 3$, the stationary flux $j(\lambda)$ may display a maximum at finite $\lambda\tau$, whose amplitude increases with $N$. The stationary flux also exhibits a minimum which is always smaller than the asymptotic value, $N/\tau_b$.

We performed a systematic study of the behavior of the exiting flux as a function of $\lambda,\tau_b$ and $N$. The flux always displays a maximum when $\tau_b$ exceeds a threshold value.   Figure \ref{fig:taubNC1} shows that the  critical value of $\tau_b$ decreases rapidly with $N$, showing that the feature observed in Fig. \ref{fig:NC1exact} is very general and occurs for smaller values of $\tau_b$ when $N$ increases.  
For $\tau_b$ below the critical value,  the stationary flux is a monotonically increasing function of $\lambda\tau$. 
The right panel of Fig. \ref{fig:taubNC1} shows the values of $\lambda\tau$ corresponding to the critical values of $\tau_b$.  

The behavior of the open probability $p_o(\lambda)$, shown in Fig. \ref{fig:NC1proba}, is consistent with Eq. (\ref{eq:jpo}). In particular one observes the appearance of a plateau whose length increases with $N$ (roughly as $N/2$). This corresponds to the situation where blockage events are rare and the output flux is close to $\lambda$. %

\subsection{Markovian versus non-Markovian models}\label{sec:stocha}
The physical assumption of constant transit and deblocking times $\tau$ and $\tau_b$, respectively, is responsible for strong memory effects  which prevent analytical solutions for general $N$ from being obtained. We therefore recently introduced Markovian models  \cite{1742-5468-2018-6-063213,0953-8984-30-30-304004}, where the average transit and deblocking times are stochastic variables given by exponential distributions of intensity $\mu$ and  $\mu^*$, respectively. The kinetic description of the Markovian model is given by a set of differential equations for the time evolution of the state probabilities $P(i,t)$ with $i\in[0\cdots N]$ giving the number of particles in the channel. Unlike the non-Markovian model, analytic solutions for the steady state properties can be obtained for arbitrary $N$ (some  generalizations of  the Markovian models for which time-dependent solutions can be obtained and could be investigated in the future \cite{Leonenko2009400,Escobar20021353}).

The channel is open for an mean time $\langle t\rangle$ and blocked for a mean time $1/\mu^*$.  The stationary flux is obtained using the previously employed recurrence arguments, giving

\begin{equation}\label{eq:j1markov}
j(\lambda)=\frac{\lambda\langle t \rangle}{\langle t \rangle+1/\mu^*}.
\end{equation}

The average time for which the Markovian system is open in a recurrence cycle is \cite{1742-5468-2018-6-063213}

\begin{equation}\label{eq:tbarmarkov}
\langle t \rangle = \frac{1}{\lambda} \sum_{j=0}^{N-1}\frac{N!}{(j+1)(N-j-1)!}  \left(\frac{\mu}{ \lambda}\right)^j.
\end{equation}

\begin{figure}
	\begin{center}
		\includegraphics[width=8cm]{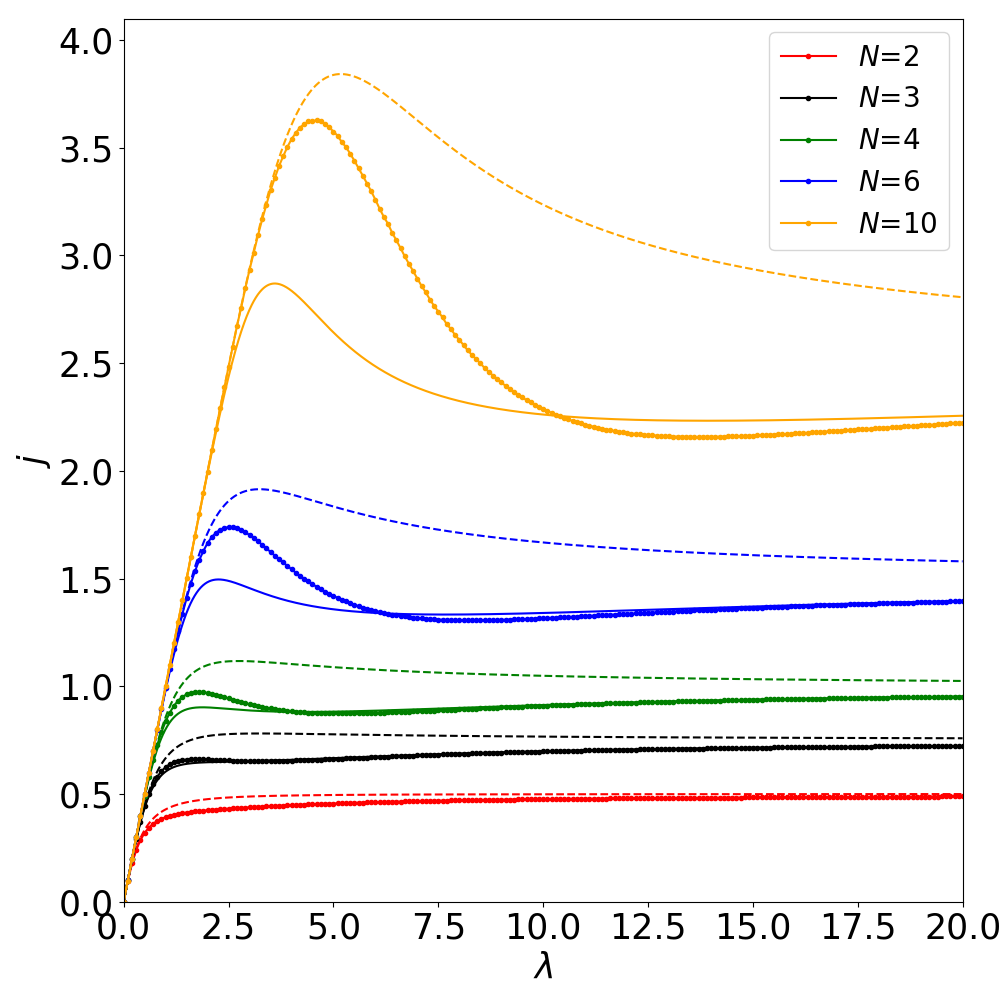}	
	\end{center}
	\caption{Comparison of stationary exiting flux $j(\lambda)$ versus $\lambda\tau$ obtained for the non-Markovian model (dotted curves), the Markovian model by setting $\mu=1/\tau$ (dashed curves) and the Markovian model by using  Eq.(\ref{eq:ansatz})  (full curves), $\tau_b = 4$.}	
	\label{fig:stochadeter}
\end{figure}

To compare the two models, $\mu$ and $\mu^*$ must be related to $\tau$, $\tau_b$ and $\lambda$. Eq. (\ref{eq:j1markov}) with Eq. (\ref{eq:j}) shows that $\mu^*$ must equal $1/\tau_b$. To obtain an expression for $\mu$, we consider the system's behavior at low and high intensity. When $\lambda\tau<<1$, the non-Markovian transit time is equal to $\tau$. The mean transit time is $1/\mu$ in the the Markovian model. A first approach is to therefore set $\mu=1/\tau$. When $\lambda \tau>>1$, we expect $\mu$ to decrease to zero. Figure \ref{fig:stochadeter} shows that the stationary flux of the  Markovian model is always larger than that of the non-Markovian model. Even though the two models behave similarly at small and large input intensity, they increasingly deviate for intermediate intensities with increasing $N$.

To obtain an exact mapping (in the steady state) we equate the mean blocking time of the two models. For $N=2$ we equate $\langle t \rangle$ given by Eq. (\ref{eq:tbarmarkov}) with the result for the non-Markovian model \cite{Gabrielli2013}, $\langle t \rangle=(2-e^{-\lambda\tau})/(\lambda(1-e^{-\lambda\tau}))$. The expressions are identical when
\begin{equation}
\mu=\frac{\lambda}{e^{\lambda\tau}-1}.
\end{equation}
With this mapping, we recover the aforementioned expected limiting behaviour for both extremes of entering flux intensity. We emphasize that the transient regimes of the two models are different (See the Appendix \ref{sec:appendix} for a similar model where time-dependent analytic solutions are obtained).

The same procedure can be carried out for $N=3$ using Eqs. (\ref{eq:temps}) and (\ref{eq:temps2}), but the resulting expression for $\mu$ is considerably  more complex. For general $N$ we therefore propose the following ansatz, taking a similar form as the mapping for $N=2$: 

\begin{equation}
\mu=\frac{2\lambda/N}{e^{2\lambda\tau/N}-1}\label{eq:ansatz}
\end{equation}
which behaves as $1/\tau$ at low intensity and approaches zero exponentially at large intensity.
Substituting  Eq.(\ref{eq:ansatz}) into Eq.(\ref{eq:j1markov}), produces a lower bound of the stationary flux (full curves). Furthermore, the maximum of the flux is underestimated and shifted to a smaller  intensity  than in the non-Markovian model. For  $N=3$ and $N=4$, the curves are very close to the results of the non-Markovian model. For $N > 4$, the ansatz leads to a significant underestimation of the exiting flux for small $\lambda \tau$. 

\section{Bundle model}
\label{sec:multimodel}
We now consider a bundle of $N_c$ identical channels. Each channel has the same properties as the single channel model above, i.e.
blockage occurs when $N$ particles are present in a channel at the same time. In the following we assume that the total intensity, $\Lambda=\lambda N_c$, is constant and is equally distributed over the open channels. Thus, after $k$ blockages the intensity on each of these open $N_c-k$ channels is 
\begin{equation}
 \lambda_k=\lambda \frac{N_c}{N_c-k}.
\end{equation}
Since a blocked channel releases all particles after finite time $\tau_b$, the system's mean output flux evolves towards a non-zero stationary value. 
The bundle has two states: {\it open} in which at least one of the constituent channels is open and {\it closed} if all the constituent channels are blocked. If a particle arrives while the bundle is in the latter state, it is rejected. 

Equations (\ref{eq:Po},\ref{eq:j},\ref{eq:mt}) cannot be applied to the channel bundle in the steady state, as it does not cycle between closed and empty states for finite intensity $\Lambda$. 
In the limit of very large intensities, however,  we have 
\begin{equation}
P_{o}(\Lambda)\sim \frac{1}{\Lambda}
\label{eq:BPo}
\end{equation}
and
\begin{equation}
\lim_{\Lambda\rightarrow\infty}\langle M \rangle =0 
\end{equation}
where $\langle M\rangle$ is the mean number of exiting particles that are not the due to blockage releases, and 
\begin{equation}
\lim_{\Lambda\rightarrow\infty}J(\Lambda)=\frac{NN_c}{\tau_b}.
\label{eq:BJ}
\end{equation}
In this limit the intensity is so high that all channels block instantaneously and simultaneously and the blocked particles 
are released after a time $\tau_b$. The exiting flux is entirely the result of these releases. 

The analogue of Eq. (\ref{eq:jpo}),
\begin{equation} 
J(\Lambda)=\Lambda P_{o}(\Lambda),
\label{eq:JPo}
\end{equation}
is valid for arbitrary intensity since, as for the single channel case, it is result of the conservation of particle number. 

\subsection{Exact solution: $N=1$}
\label{sec:BN1}
When $N>1$ a particle may traverse the channel in a time $\tau$ without causing a blockage. In comparison, the $N=1$ model is singular as  no unimpeded transit is possible: each entering particle causes a blockage that lasts for a fixed time, $\tau_b$. The variable $\tau$ is thus absent in this model.

Despite the relative simplicity of the $N=1$ model, its dynamics cannot be written as a system of differential equations for the state probabilities $P(i,t)$, where $i$ denotes the number of blocked channels at time $t$ (in Appendix \ref{sec:appendix} the full time dependent solution for $N_c=2$ is presented). 
However, in the stationary state, by applying detailed balance (known as the ``rate up - rate down" principle in queuing theory), one has
\begin{equation}
\Lambda P(i)=(i+1)\frac{P(i+1)}{\tau_b}
\end{equation}
Solving the difference equation and applying conservation of the total probability leads to

\begin{equation}
P(k)=\frac{(\Lambda\tau_b)^k}{k!\sum_{n=0}^{N_c}\frac{(\Lambda\tau_b)^n}{n!}},\;k\in [0,N_c]
\end{equation}
The stationary exiting flux is given by Eq. (\ref{eq:JPo}) with
\begin{equation}\label{eq:PoexactN1}
P_o=1-P(N_c)=1-\frac{(\Lambda\tau_b)^{N_c}}{N_c!\sum_{n=0}^{N_c}\frac{(\Lambda\tau_b)^n}{n!}}
\end{equation}
The result can be written in the form
\begin{equation}\label{eq:DJ1}
 J(\Lambda)=\Lambda\left(1-\frac{(\Lambda\tau_b)^{N_c}}{e^{\Lambda\tau_b}\Gamma(1+N_c,\Lambda\tau_b)}\right),
\end{equation}
where $\Gamma(n,x)$ is the incomplete gamma function.
The asymptotic behavior at small intensity $\Lambda$ is
\begin{equation}
 J=\Lambda\left[1-\frac{(\Lambda\tau_b)^{N_c}}{N_c!}+O((\Lambda\tau_b)^{N_c+1})\right],
\end{equation}
whereas at large intensity the flux behaves as,
\begin{equation}
 J=\frac{N_c}{\tau_b}\left[1-\frac{1}{\Lambda\tau_b}+O\left(\frac{1}{(\Lambda\tau_b)^2}\right)\right], 
\end{equation}
whose leading term is in accordance with Eq. (\ref{eq:BJ}). For all values of $N_c$, $J(\Lambda)$ is always a monotonically increasing function of $\Lambda$. 

We note that the expression for $P(N_c)$ is Erlang's first formula \cite{takacs1969,Medhi:1991:SMQ:113159} for a stochastic queuing process
with $N_c$ servers with exponential entry and service time distributions under the condition that when all servers are busy an arrival is rejected. Both models are `birth-and-death' processes that have the {\it same} stationary solution. Their transient regimes, however, are significantly different. See Appendix \ref{sec:appendix}.

\begin{figure}[t]
	\begin{center}
		\includegraphics[width=8cm]{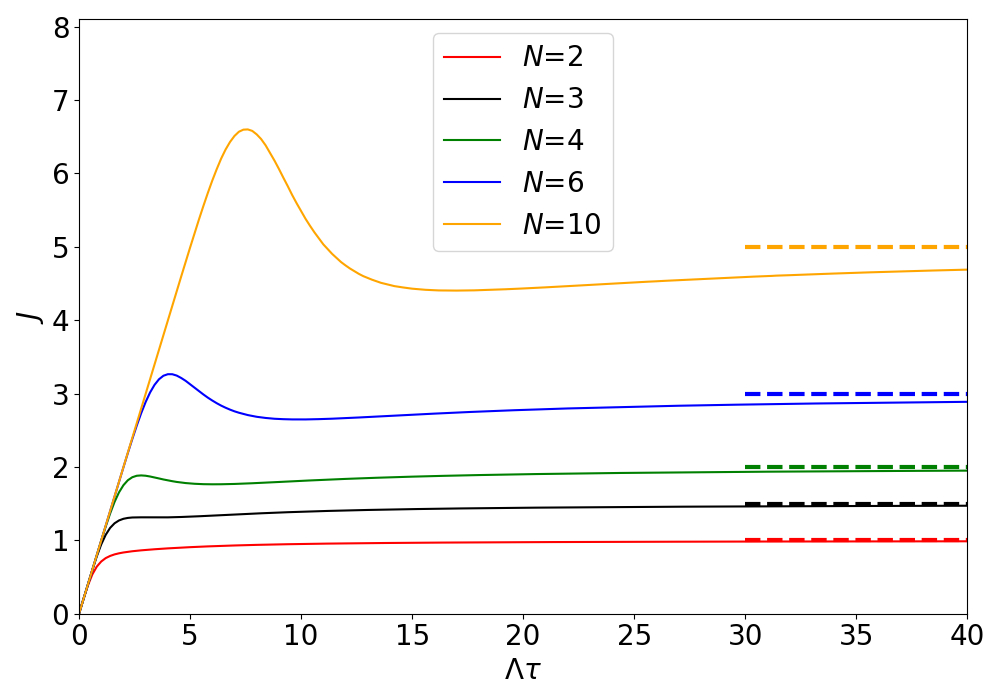}\\
				\includegraphics[width=8cm]{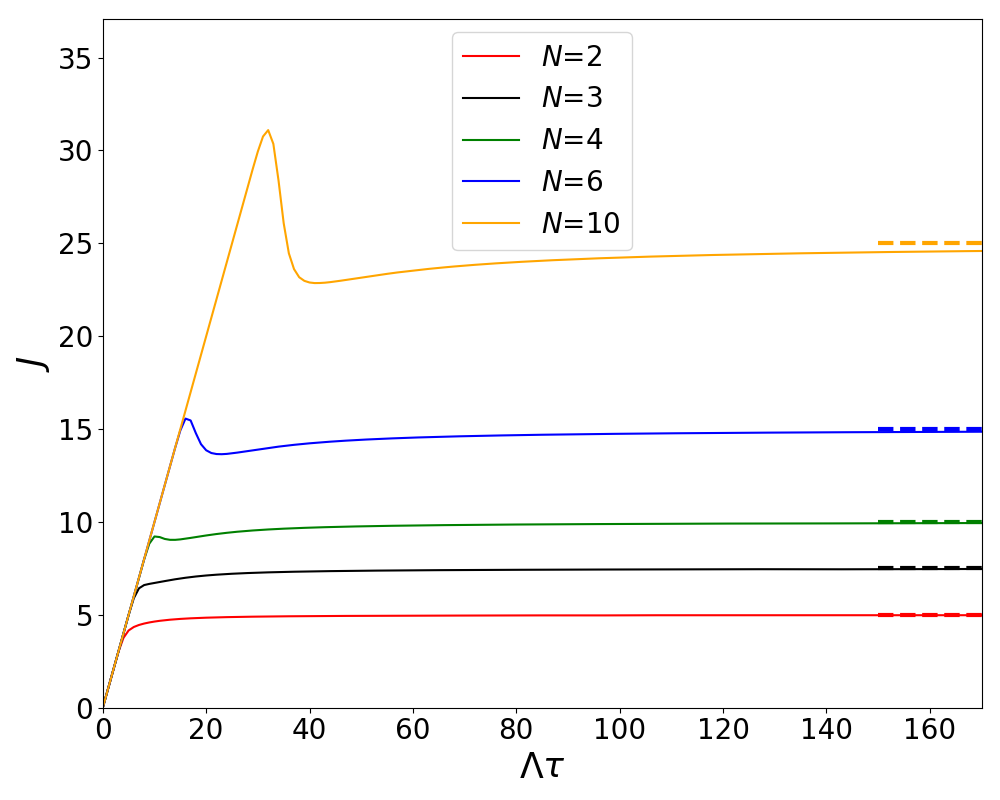}
	\end{center}
	\caption{  $J(\Lambda)$ versus $\Lambda\tau$ for $N=2,3,4,6,10$ and $\tau_b=4$, for (top) $N_c=2$, and   (bottom) $N_c=10$. The dashed lines correspond to the asymptotic values, Eq. (\ref{eq:BJ}).}
	\label{fig:NCflux}
\end{figure}

\subsection{Simulation results: $N>1$}
For the multichannel  models, no exact solution can be obtained for $N>1$. Therefore we have performed numerical simulations to obtain 
the stationary exiting flux, $J$ and the stationary probability that at least one channel is open, $P_o$ for bundles composed of different numbers of channels with increasing capacity $N$ and for $\tau_b=4\tau$. All quantities were investigated as a function of the mean incoming flux $\Lambda$.  As discussed in the previous  section, the stationary flux rapidly displays a maximum at a finite value of $\lambda \tau$ when $N>1$.

Figure \ref{fig:NCflux} shows $J$  as a function of $\Lambda\tau$ for  $N_c=2,10$ and for $N=2,3,4,6,10$.
When $\Lambda\tau\ll 1$, the rate of incoming particles is very small and the finite capacity of the channel is rarely reached, meaning that blockage events are scarce. The stationary exiting flux is therefore equal to the input flux, $J\simeq \Lambda$. This behavior is observed for a larger range of $\Lambda\tau$ for larger values of $N$ and $N_c$ .  


\begin{figure}[th]
	\begin{center}
		\includegraphics[width=7.6cm]{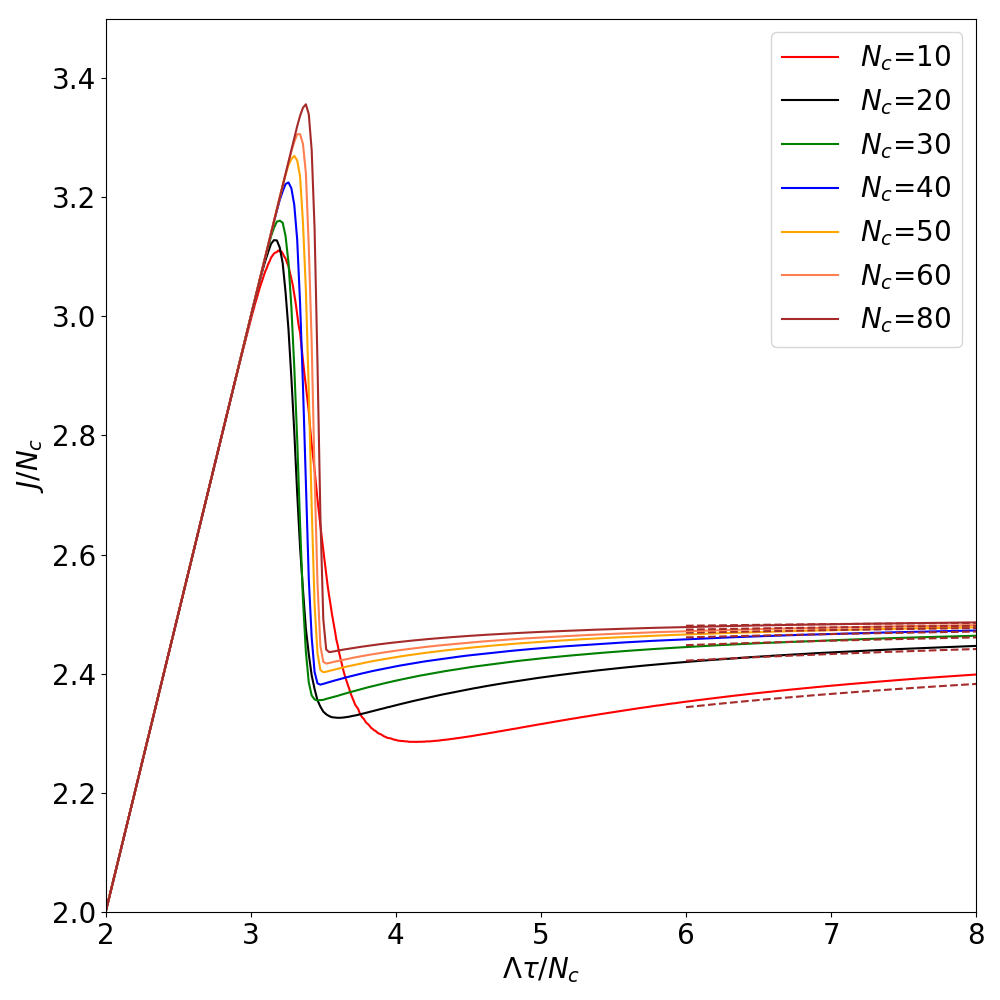}	
	\end{center}
	\caption{Exiting flux of a channel bundle composed of $N_c$ channels each with $N=10$:   $J/N_c$ versus $\Lambda\tau/N_c$ for different values of $N_c$ and $\tau_b=4$. The dashed curves show the asymptotic behavior.}
	\label{fig:Nasymp}
\end{figure}

Figure \ref{fig:Nasymp} shows the rescaled flux $J/N_c$ versus the rescaled intensity $\Lambda/N_c$ for different values of $N_c$. In the low intensity regime $J$ is equal to $\Lambda$ and displays a finite discontinuity at $\Lambda_c$. At high intensities the curve evolves towards an asymptote and is quite well described by
\begin{equation}\label{eq:fit}
J\simeq \frac{N_c N}{\tau_b}\left(1-\frac{3N}{2\Lambda\tau_b}\right).
\end{equation}
We observe an abrupt change of kinetic behavior: Below the critical value $\Lambda_c$, almost all particles cross the bundle without triggering a significant number of blockages, whereas for larger $\Lambda$, all channels are closed and the stationary flux is essentially given by the release of blocked particles. 

\begin{figure}[th]
	\begin{center}
		\includegraphics[width=8cm]{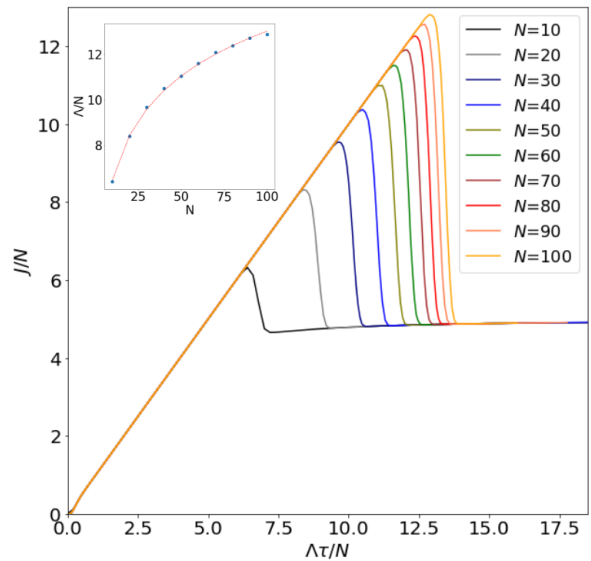}	
	\end{center}
	\caption{$N_c=20$:   $J/N$ versus $\Lambda\tau/N$ for different values of $N$ and $\tau_b=4$. The dashed curves correspond to the asymptotic behaviors. The inset shows the value of $\Lambda$ that maximizes the exiting flux as a function of $N$ and the red dashed line shows the logarithmic fit, of form $2.73\log(0.89N)+0.65$.}
	\label{fig:NC10asymp}
\end{figure}

Fig. \ref{fig:NC10asymp} shows the rescaled flux $J/N$ versus the rescaled intensity $\Lambda/N$ for different values of $N$ for a given $N_c=10$. At a low input intensity, the exiting flux $J$ is equal to $\Lambda$ until it reaches a maximum close to a critical value that closely follows a logarithmic law, as shown in the inset of Fig. \ref{fig:NC10asymp}. 
For higher values of input intensity, the rescaled exiting flux, for all values of $N$, rapidly collapses to a single curve, whose best fit is again given by Eq. (\ref{eq:fit}). 

\section{Flux optimization}\label{sec:compare}

\begin{figure}
	
	\includegraphics[width=8.0cm]{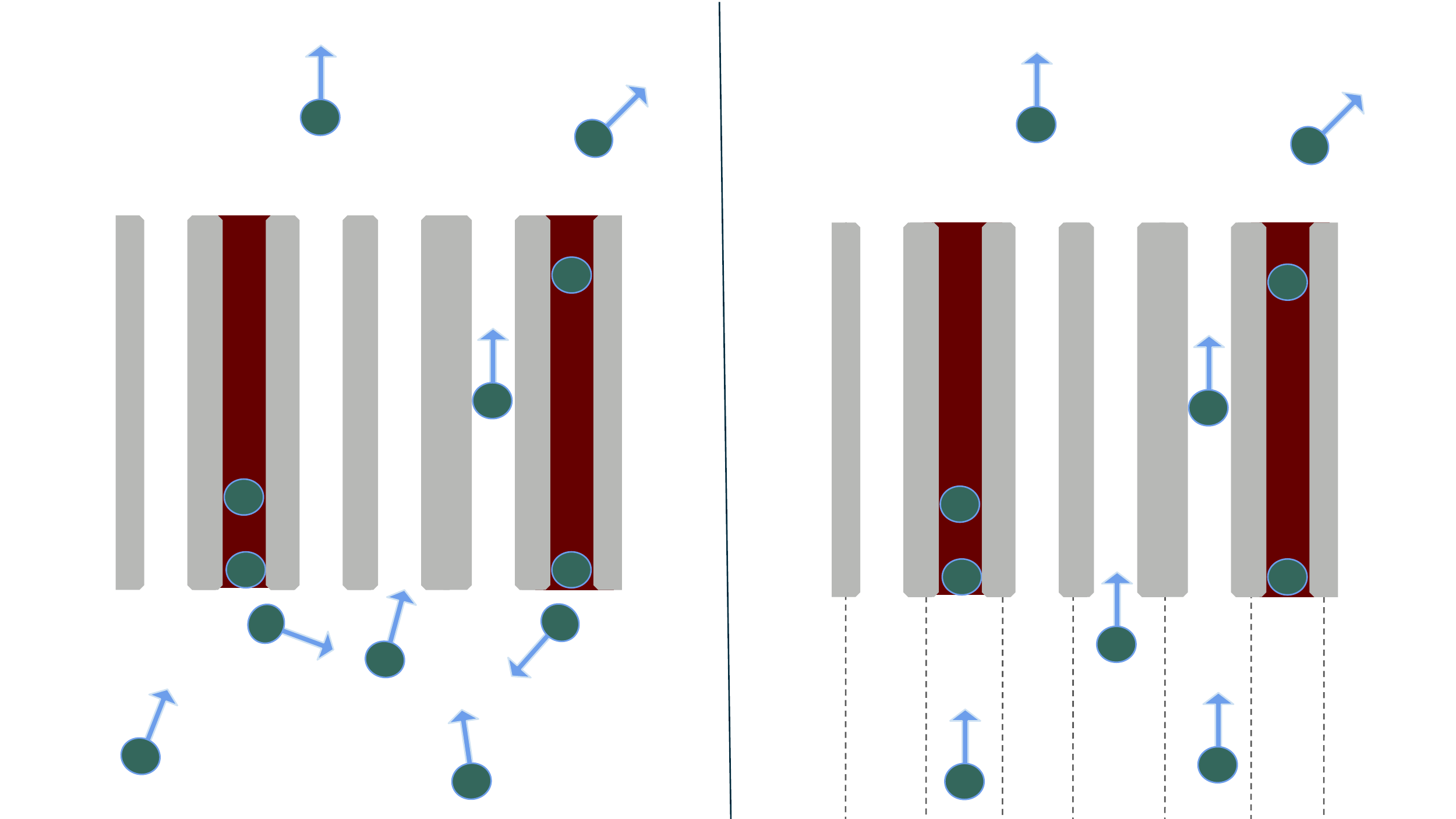}
	
	\caption{Left: A bundle composed of $N_c=4$ coupled channels, each of capacity $N=2$, sharing an incoming flux of intensity $\Lambda$. Right: Four uncoupled channels, each of capacity $N=2$, each receiving an incoming flux of intensity $\Lambda/4$.
}\label{fig:config1}
\end{figure}

\begin{figure*}

		\includegraphics[width=17cm]{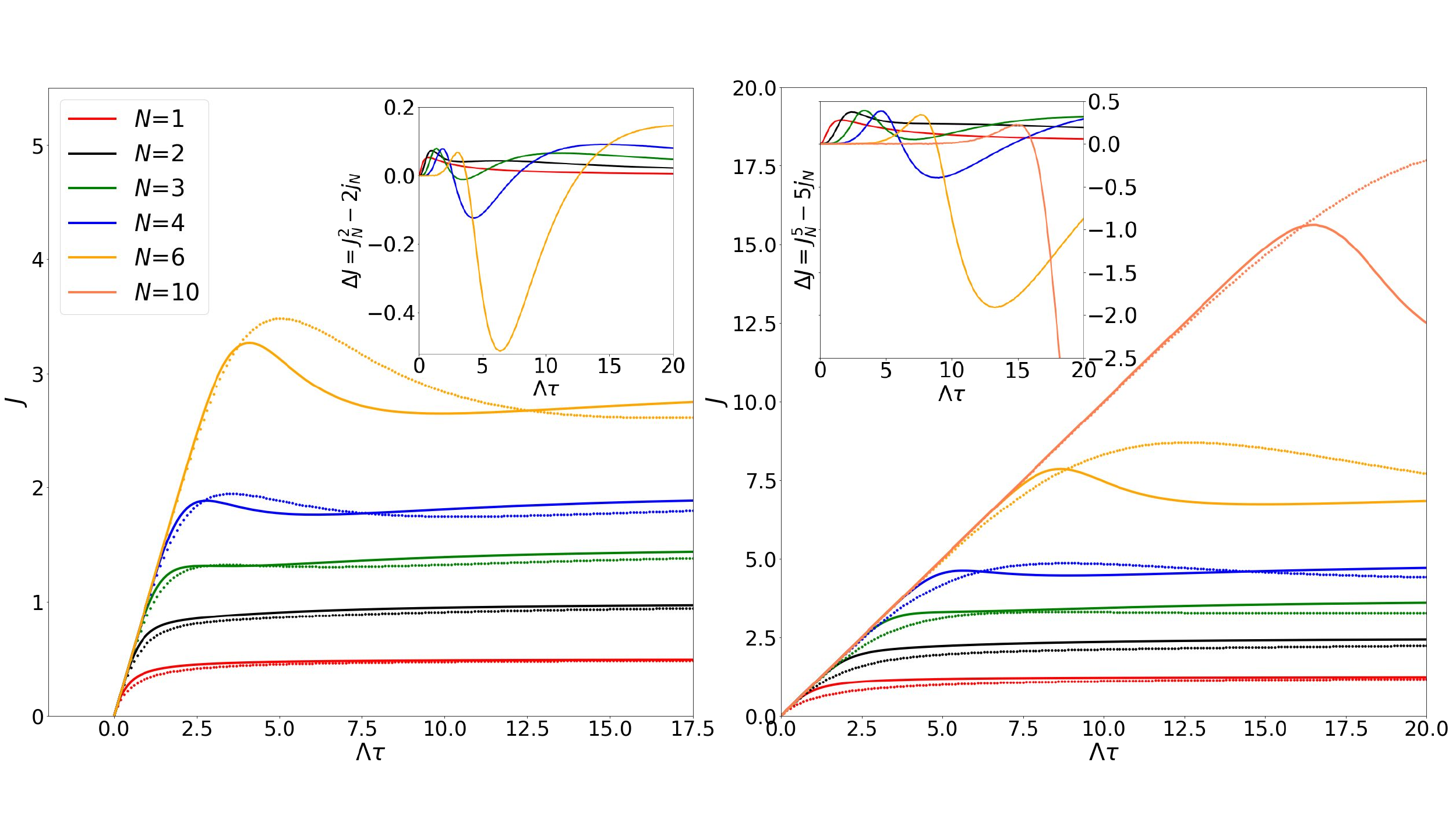}

	\caption{ Total exiting flux from the coupled (lines) and uncoupled (dotted) channel configurations. From Left to right, $N_c = 2,5$, $\tau_b = 4$. The insets show the difference between the two.}
	\label{fig:CoupleUncouple}
\end{figure*}

Here we compare the stationary flux of a bundle of coupled channels with that of a bundle of uncoupled channels and one high capacity (HC) channel.  The transport efficiency is measured by the difference in output flux, $\Delta J$. The systems are chosen so that in the limits of low and high input flux intensity, $\Delta J = 0$. In the low intensity limit, since blocking events are rare, the exiting flux is equal the input flux $\Lambda$, irrespective of the configuration. In the high intensity limit, Eqs.(\ref{eq:Poh}) and (\ref{eq:BJ}) demonstrate that the exiting fluxes of the single high capacity, bundled uncoupled or coupled channels are also be equal.  Since the stationary flux of a bundle of coupled channels displays non-trivial behavior with increasing $N_c$ and $N$, we therefore expect non-trivial behavior of $\Delta J$.

\subsection{Coupled versus uncoupled channels}

\begin{figure}[h]

	\includegraphics[width=8cm]{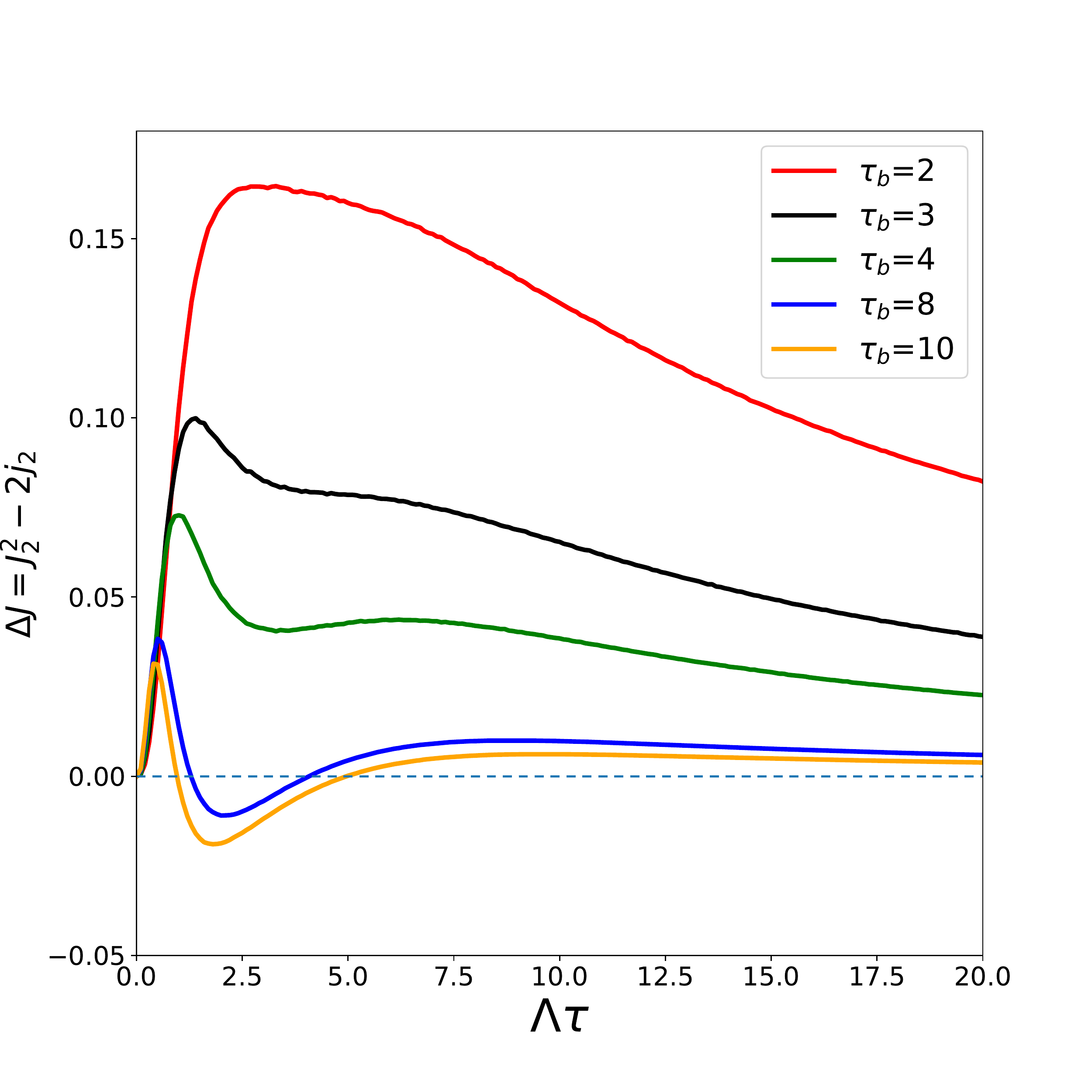}
	
	\caption{Difference in the output flux between $N_c=2$ coupled and uncoupled channels, with $N=2$ as a function of intensity $\Lambda$, for different values of $\tau_b$.}

	\label{fig:LCLCtaub}
\end{figure}

\begin{figure}[h]
	
	\includegraphics[width=8.0cm]{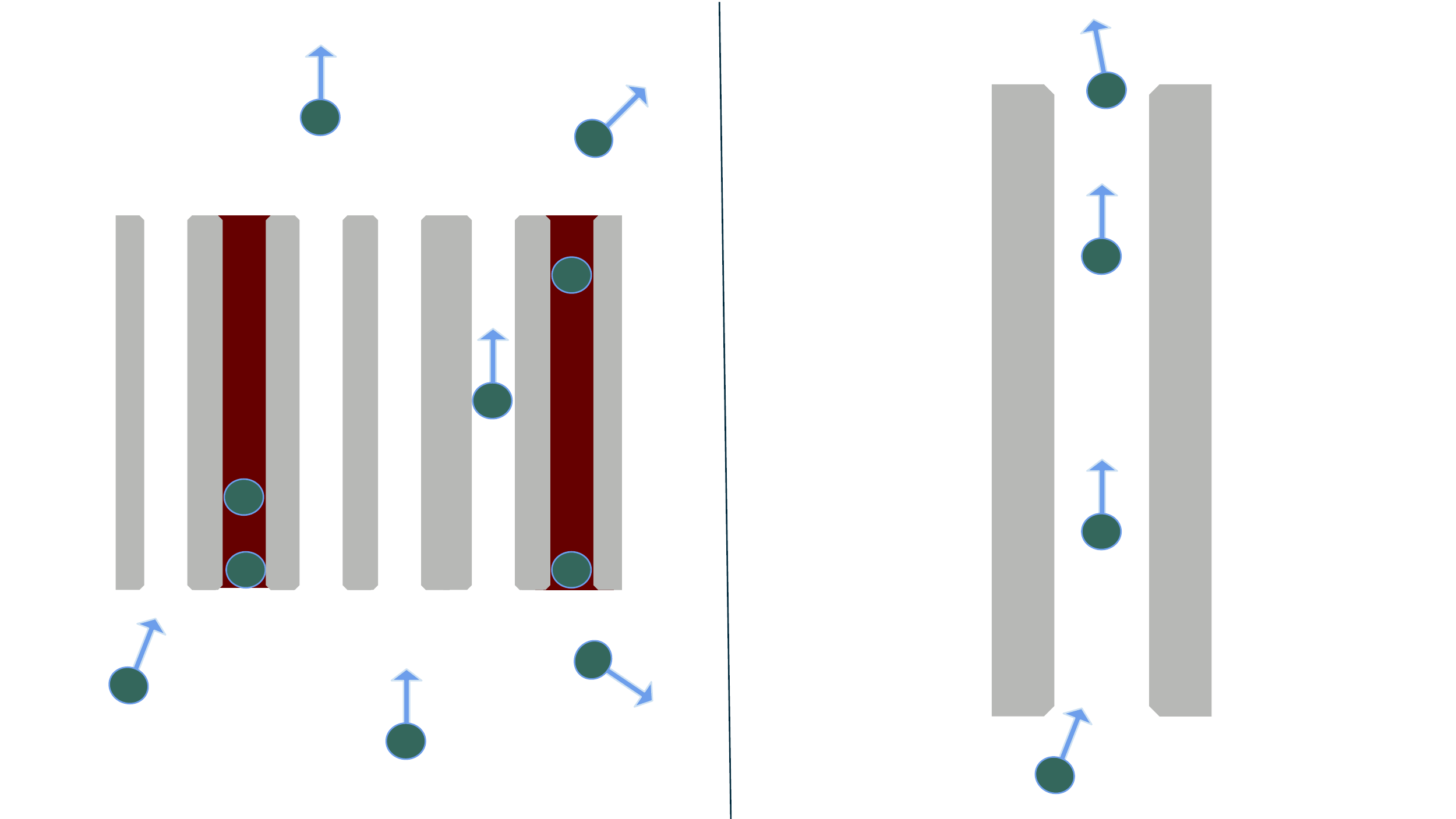}
	
	\caption{Left: A bundle composed of $N_c=4$ channels, each with capacity $N=2$, sharing an incoming particle flux of intensity $\Lambda$. Right: A single high capacity channel of capacity $NN_c=8$.}\label{fig:config2}
\end{figure}

We first compare a channel bundle composed of $N_c$ channels, each of capacity $N$. The entering flux, $\Lambda$, is equally distributed over the coupled open channels. In contrast, the $N_c$ independent channels, each of capacity $N$, each receive an incoming flux of intensity $\Lambda/N_c$. Fig. \ref{fig:config1} illustrates the configurations compared.  The difference in the output flux of the two configurations is defined as:
\begin{equation}\label{eq:CoupUncoup}
	\Delta J = J^{N_c}_{N} (\Lambda) - N_c j_{N} ( \Lambda / N_c)  .
\end{equation}

Simulation results comparing the output flux of each configuration for $N_c= 2,5$, $\tau_b = 4$, are shown in Fig. \ref{fig:CoupleUncouple}. The differences are shown in the inset of the figure.  At low intensity, for all configurations, the output flux is approximately equal for each setup, and linearly increases with $\Lambda$ until a critical value, which itself is a monotonically increasing function of $N$.  

The behavior for $N=1$ can be understood quantitatively using the results of Sec. \ref{sec:BN1} and Eq. (\ref{eq:jolo}). At low intensity, the flux difference is
\begin{equation}
 \Delta J =\frac{\Lambda^2\tau_b}{N_c}+O(\Lambda^3),
\end{equation}
and at high density we find
\begin{equation}
 \Delta J \simeq\frac{N_c(N_c-1)}{\Lambda\tau_b^2}
\end{equation}
and one can confirm that $\Delta J>0$ for $0<\Lambda<\infty$. We conclude that the coupled channels are always more efficient than the uncoupled ones. The difference is maximized for a finite value of $\Lambda$.

For all $N>1$, we note the appearance of two maxima in the flux difference with an intervening minimum. The increased complexity is due to the presence of two characteristic times, the transit time $\tau$ and the blockage time $\tau_b$ (while the $N=1$ system has only the latter).

Fig. \ref{fig:LCLCtaub} shows the flux difference between the bundle configurations at $Nc=2$, $N=2$, as a function of intensity of entering flux, $\Lambda$, for different values of $\tau_b$. For $\tau_b > 2$ the behavior is more complex after the first maximum, with the appearance of a minimum followed by a second maximum before tending towards zero.

\subsection{Single HC channel verses coupled LC channels }

\begin{figure}
	\begin{center}
		\includegraphics[width=8cm]{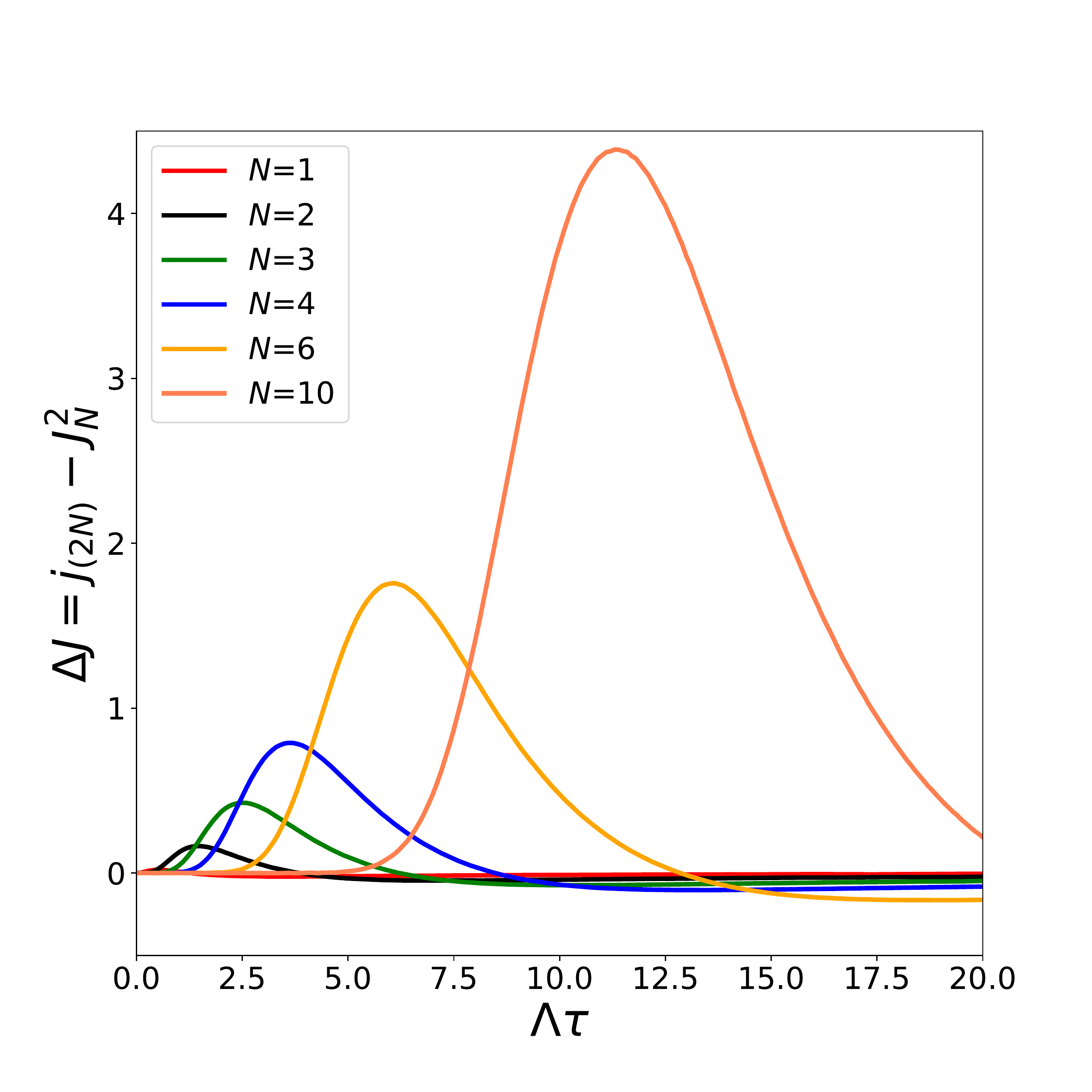}\\
		\includegraphics[width=8cm]{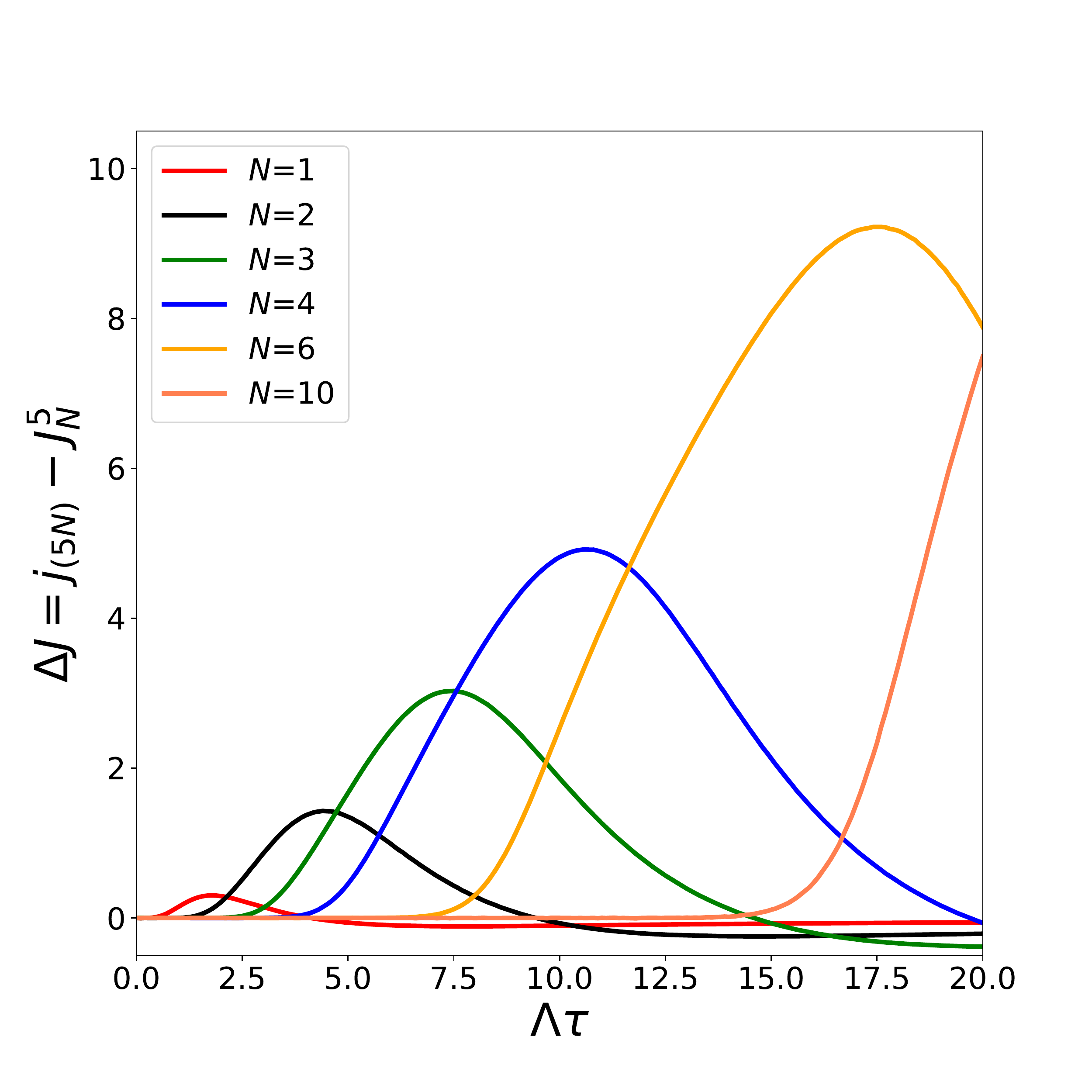}
	\end{center}
	\caption{Single HC channel, versus coupled LC channels. From top to bottom, $N_c = 2,5$. All curves display a single maximum followed by a single minimum, before tending to zero at high intensity.  
	 }
	\label{fig:HCLCcouple}
\end{figure}
The transport efficiency of a single 'high capacity' (HC) channel, with a bundle of several coupled channels, of a proportionately reduced capacity, is now compared. The two configurations are illustrated in Fig. \ref{fig:config2}. Figure \ref{fig:HCLCcouple} shows the difference in the output stationary state flux between a single HC channel and a bundle of $N_c= 2,5$ coupled LC channels, with different capacities. The flux difference in this case is

\begin{equation}
	\Delta J = j_{N_c N} (\Lambda) - J^{N_c}_{N} ( \Lambda)  .
\end{equation} 

For all $N$, the flux difference displays a minimum, followed by a maximum, before tending towards zero for increasing intensity. The amplitudes of the maxima are always greater than those of the minima.

\section{Discussion}\label{sec:discussion}
We have presented a model of blockage in channel bundles that may be relevant for various physical processes. The bundle consists of $N_c$ constituent channels, each with a capacity of $N$. A particle transits through an open channel in time $\tau$, but if $N$ particles are simultaneously present in a channel, it is blocked for a time $\tau_b$, before being emptied. While blocked, the entering flux is redistributed over the remaining open channels. A  bundle of channels is open if at least one of its constituent channels is not blocked. If the entering stream is of constant intensity the bundle evolves to a stationary state with a steady exiting flux, or throughput, that depends on the intensity, $\tau_b$ and $N$. In the steady state the exiting flux is simply related to the probability that the bundle is open. For a single channel with capacity $N>1$ the exiting flux displays a maximum value at finite intensity if $\tau_b$ is sufficiently large. If not, the exiting flux increases monotonically with the intensity. A Markovian model with stochastic transit and blockage times, inspired by queuing theory, can be made to display the same steady state behavior with an appropriate mapping between the two models' parameters. The transient behavior is, however, quite different. This suggests that, in the steady state, the details of the transport mechanisms and the triggered releases are not important. For large $N$, the models display an abrupt change from a state with few blockages to one in which the bundle is permanently blocked and the output flux is entirely due to the release of blocked particles. This behavior raises new questions about whether more general relationships describing the abrupt transitions in dynamics may be obtained for general $N$ and $N_c$. The transport efficiency of a bundle in which the entering flux is equally distributed over the open channels was also compared with a bundle composed of independent channels. For $N=1$, the coupled channels always have a higher throughput, but for larger values of $N$ the behavior is more complex. 

\acknowledgments

We thank Jacques Resing, Karim Guerouate et Valentin Wiener for useful discussions.

\appendix

\section{Full Solution of the Channel Bundle with $N_c=2$, $N=1$}\label{sec:appendix}

This model can be solved exactly by using an approach similar to that used to solve the single channel model with $N=3$ \cite{barre2015a}. This requires introducing the partial probabilities $Q(n,t)$ that at time $t$, $n$ particles have entered such that at least one channel is open. The open probability of the bundle is then given by
\begin{equation}
 P_o(t)=\sum_{n\ge 0}Q(n,t)
\end{equation}
Since blockage is not possible for $n=0,1$ the first two probabilities are trivial:
\begin{align}
	Q(0,t)&=e^{-\Lambda t}\, ,\\
	Q(1,t)&=\Lambda t e^{-\Lambda t}\, ,
\end{align}
For  $n\geq 2$, the probability $Q(n,t)$ can be written as:
\begin{align}\label{eq:qn}
	Q(n,t)&= \int_0^\infty \prod_{i=0}^{n-1} dt_i \Lambda e^{-\Lambda t_i} \int_0^\infty dt' e^{-\Lambda t'}\delta(t-\sum_{i=0}^{n-1} t_i -t')\nonumber\\
	&\times\prod_{j=1}^{n-2}\theta(t_j+t_{j+1}-\tau_b)
	\nonumber\\&\prod_{k=1}^{n-1}(\theta(\tau_b-t_k) e^{\Lambda (\tau_b-t_k)}+\theta(t_k-\tau_b))
	\nonumber\\&\times\theta(t'+t_{n-1}-\tau_b)\, .
\end{align}
The $n$-fold integral corresponds to all events  of $n$ incoming articles at the entry of the channel
with the associated constraint ($\delta$ function) which  occurs between $0$ and $t$. The first product of  Heaviside functions expresses the fact that a new particle can enter if both channels are not blocked, which imposes   the condition that the time since the entry of the second last  particle is larger than  $\tau_b$; The last product corresponds to both complementary situations: either the last entered particle leads to a complete blockage of two channels (The exponential factor expresses that no particle can enter in a duration $\tau-t_k$) or only one channel is blocked by the last entered particle. The last Heaviside function imposes the requirement that at least one channel be open at time $t$.

Taking the Laplace transform of $Q(n,t)$, the integral over $t_0$ is trivial and the integral over $t'$ must be split into two parts, which finally gives 
\begin{align}\label{eq:qnu}
	\tilde{Q}(n,u)&=\left(\frac{\Lambda  }{\Lambda+u }\right)^2\left[e^{-u\tau_b} \int_0^{\tau_b}dt e^{-\Lambda t }r(n-1,u,t)\right.\nonumber\\
	&\left.+\int_{\tau_b}^\infty dt e^{-(\Lambda+u) t }r(n-1,u,t)\right]\, ,
\end{align}
with  $t_{n-1}\equiv t$.
The function $r(n-1,t,u)$ is defined as 
\begin{align}
	r(n-1,t,u)=&\int_0^\infty \prod_{i=1}^{n-2} dt_i \Lambda e^{-(\Lambda+u) t_i}  \prod_{j=1}^{n-2}\theta(t_j+t_{j+1}-\tau_b)\nonumber\\
	&\times \prod_{k=1}^{n-2}(\theta(\tau_b-t_k) e^{\Lambda (\tau_b-t_k)}+\theta(t_k-\tau_b))\, ,\label{eq:rnut}
\end{align}
By using Eq.(\ref{eq:rnut}), one infers the recurrence  relation between   $r(n,u,t)$  and $r(n-1,u,t)$
\begin{align}
	r(n,t,u)&=\Lambda\int_{\max(\tau_b-t,0)}^{\tau_b} dt'  e^{-(u+2\Lambda) t'}  e^{\Lambda \tau_b} r(n-1,t',u)\nonumber\\
	& +\Lambda \int_{\tau_b}^\infty dt'e^{-(\Lambda+u) t'}  r(n-1,t',u)\, ,\label{eq:recu}
\end{align}
with the initial condition $r(1,t,u)=1$

We define the generating function $G(z,t,u)$
\begin{equation}
G(z,t,u)=\sum_{n=1}^\infty z^{n-1} r(n,t,u)\, .
\end{equation}
Multiplying  Eq. (\ref{eq:recu}) by $z^n$ and summing over $n$, gives the  generating function $G(z,t,u)$ obeys to a similar equation
\begin{align}\label{eq:GN1N_c}
	G(z,t,u)=&1+z\int_{\max(\tau_b-t,0)}^{\tau_b} dt' \Lambda e^{-(u+2\Lambda) t'}  e^{\Lambda \tau_b} G(z,t',u)
	\nonumber\\
	& +z\int_{\tau_b}^\infty dt'\Lambda e^{-(\Lambda+u) t'}  G(z,t',u)\, .
\end{align}

For $t\geq \tau_b$, the right-hand side of Eq.(\ref{eq:GN1N_c}) is independent of $t$, which gives that
$G(z,t,u)$  is then independent of $t$. For $t<\tau_b$, by taking the two first  derivatives of Eq.\ref{eq:GN1N_c} with respect to $t$, one can rewrite the integral equation,  Eq.(\ref{eq:GN1N_c}) as an ordinary differential equation
\begin{equation}
\frac{\partial^2 G(z,t,u)}{\partial t^2}   -(2\Lambda +u)  \frac{\partial G(z,t,u)}{\partial t} +z^2\Lambda^2 e^{-u\tau_b}G(z,t,u)=0\, .
\end{equation}
whose solution is given by 
\begin{equation}\label{eq:solG}
G(z,t,u)=A(u) e^{r_1 t}+B(u)e^{r_2 t}\,
\end{equation}
where $r_1$ and $r_2$ are the solutions of the characteristic equation
\begin{equation}
r_{1,2}=\frac{2\Lambda +u \pm \sqrt{(2\Lambda +u)^2-(2z\Lambda)^2e^{-u\tau_b}}}{2}\, .
\end{equation}
The functions $A(u)$ and $B(u)$ are determined by using the boundary conditions. At $t=0$,
by using Eq.~(\ref{eq:GN1N_c}),  $G(z,0,u)$  is expressed as
\begin{align}  
	G(z,0,u)&=1+ zG(z,\tau_b,u)\frac{\Lambda}{\Lambda+u} e^{-(\Lambda+u) \tau_b}\, .
\end{align}
which gives 
\begin{align}
	A(u)\left(1-\frac{z\Lambda e^{(r_1-(\Lambda+u))\tau_b}}{\Lambda+u}\right)+\nonumber\\
	+B(u)\left(1-\frac{z\Lambda e^{(r_2-(\Lambda+u))\tau_b}}{\Lambda+u}\right)=1
\end{align}
Now, by using the first derivative of Eq.~(\ref{eq:GN1N_c}) at $t=\tau_b$,  one obtains the second boundary equation
\begin{align} 
	\frac{\partial G(z,t,u)}{\partial t}|_{t=\tau_b}= z\Lambda e^{\Lambda\tau} G(z,0,u)\, .
\end{align}
and by using Eq.(\ref{eq:solG}), one obtains
\begin{align}
	A(u)\left(r_1 e^{(r_1-\Lambda)\tau_b} -z\Lambda \right)+B(u)\left(r_2 e^{(r_2-\Lambda)\tau_b} -z\Lambda \right)=0
\end{align}
In the Laplace space,  the open probability $\tilde{P}_o(u)$ is given by 
\begin{align}
	\tilde{P}_o(u)&=\sum_{n\ge 0}\tilde{Q}(n,u)\nonumber\\
	&=\tilde{Q}(0,u)+\tilde{Q}(1,u)+
	e^{-u\tau_b}\left(\frac{\Lambda  }{\Lambda+u }\right)^2\times\nonumber\\
	&\times\left(\int_{0}^{\tau_b}dt e^{-\Lambda t}G(1,t,u)+\frac{ e^{-\Lambda \tau_b}G(1,\tau_b,u)}{\Lambda+u}\right)
\end{align}
By using Eq.(\ref{eq:qnu}) and Eq.(\ref{eq:solG}), this can be expressed as
\begin{align}
	\tilde{P}_o(u)&=e^{-u\tau_b}\left(\frac{\Lambda  }{\Lambda+u }\right)^2\nonumber\\
	&\left[A(u)\left(\frac{1-e^{(r_1-\Lambda)\tau_b}}{\Lambda-r_1}
	+\frac{e^{(r_1-\Lambda)\tau_b}}{\Lambda+u}\right)+\right.\nonumber\\
	&+\left. B(u)\left(\frac{1-e^{(r_2-\Lambda)\tau_b }}{\Lambda-r_2}
	+ \frac{e^{(r_2-\Lambda)\tau_b}}{\Lambda+u} \right)\right]
	\nonumber\\
	&+\frac{1}{\Lambda+u}+\frac{\Lambda}{(\Lambda+u)^2}\, .\label{eq:poufinal}
\end{align}
 The functions $A(u)$ and  $B(u)$ can be determined by using the boundary conditions, but the lengthy expressions are not  displayed here.
 The output flux $J(t)$ is entirely due to the release of blocked particles and is therefore given by
 \begin{equation}
  J(t)=\Lambda P_o(t-\tau_b)
 \end{equation}

The stationary open probability is 
$P_o(\infty)= \lim_{u\rightarrow 0}  u\tilde{P}_o(u)$.
By using  Eq.(\ref{eq:poufinal}), the stationary open probability $P_o$ is then given by
\begin{equation}
P_o=\left(\tau_b+\frac{1}{\Lambda}\right)\lim_{u\rightarrow 0}[u(A(u)+B(u))]
\end{equation}
After a tedious but straightforward calculation, one finally obtains
\begin{equation}
P_o(\infty)=\frac{1+\Lambda\tau_b}{1+\Lambda\tau_b+(\Lambda\tau_b)^2/2}\,
\end{equation}
which corresponds to the Erlang formula, Eq. (\ref{eq:PoexactN1}).

 Let us now compare with the stochastic model with $N_c=2$ and $N=1$ with parameters $\Lambda$ and $\mu^*$. 
	The state of the system is defined by the probabilities $P_0(t)$, $P_2(t)$, $P_2(t)$ of having zero, one and two particles in the bundle at time $t$, respectively. They evolve according to the coupled differential equations
	\begin{align}
	\frac{dP_0(t)}{dt}&=-\Lambda P_0(t)+\mu^*P_1(t)\nonumber\\
	\frac{dP_1(t)}{dt}&=\Lambda P_0(t)-(\Lambda+\mu^*) P_1(t)+2\mu^*P_2(t)\nonumber\\
	\frac{dP_2(t)}{dt}&=\Lambda P_1(t)-2\mu^*P_2(t)
	\end{align}

\begin{figure}
	\begin{center}
		\includegraphics[width=8cm]{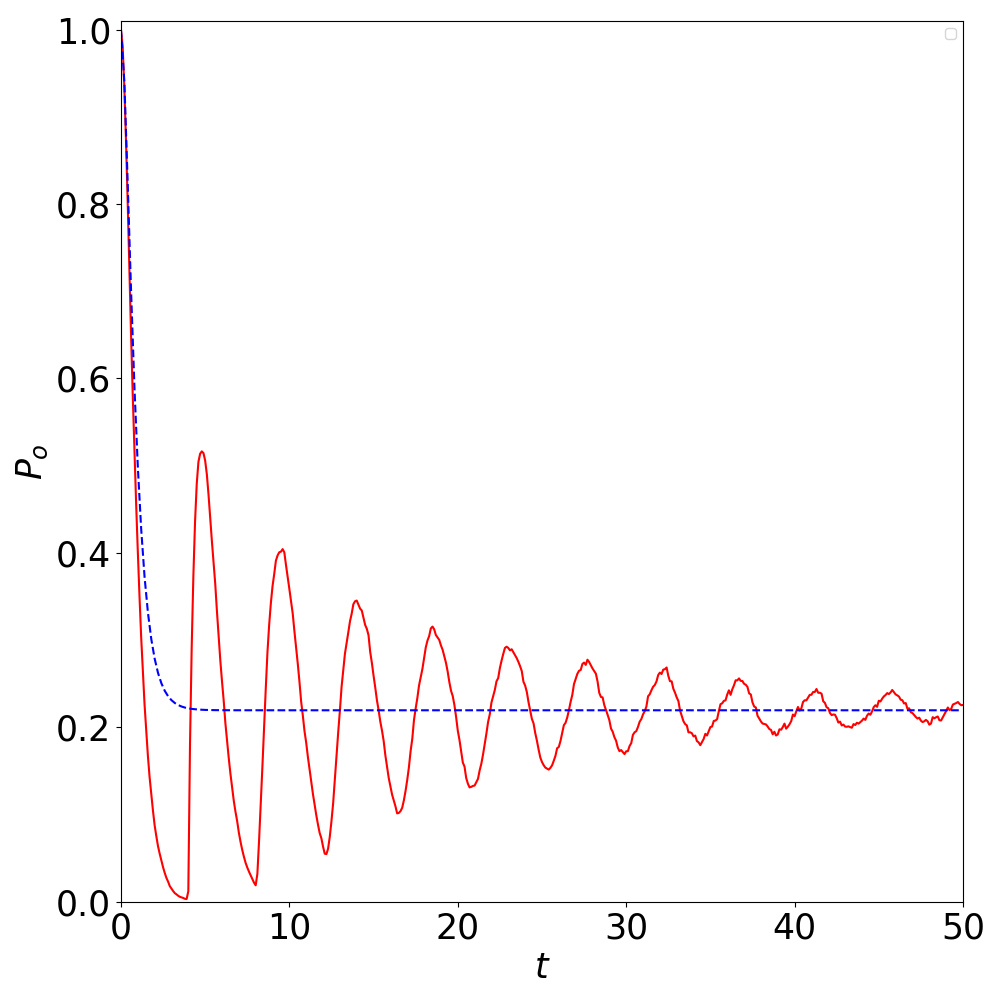}	
	\end{center}
	\caption{Time evolution of the open probability $P_o(t)$ for the non-Markovian model (full curves) and the Markovian model for $N_c=2$ and $N=1$ with $\Lambda=2$ and $\tau_b=4$ }	
	\label{fig:stochadeterflux}
\end{figure}
Taking the Laplace transform, and solving the linear system of algebraic equations, one obtains the Laplace transform of the open probability, $\tilde{P}_o(u)$
	\begin{align}
	\tilde{P}_o(u)&=\tilde{P}_0(u)+\tilde{P}_1(u)\nonumber\\
	&=\frac{(u+\mu^*)(u+2\mu^*+2\Lambda)}{u(u^2+(2\Lambda+3\mu^*)u+2\mu^{*2}+2\mu^*\Lambda+\Lambda^2)}
	     \end{align}

 Finally, calculating the inverse Laplace transform and substituting $\mu^*=1/\tau_b$, one obtains
	\begin{align}
P_o(t)&=P_o(\infty)+\frac{(\Lambda\tau_b)^2 e^{-\Lambda(2+\frac{3}{\Lambda\tau_b})t}}{1+2\Lambda\tau_b+2\Lambda^2\tau_b^2}\left[\cosh\left(\frac{t}{\tau_b'}\right)+\right.\nonumber\\
&+\left.\left(2\Lambda\tau_b'+3\frac{\tau_b'}{\tau_b}\right)\sinh\left(\frac{t}{\tau_b'}\right)\right]
  \end{align}
  with $\tau_b'=\frac{\tau_b}{\sqrt{1+4\Lambda\tau_b}}$.

Figure \ref{fig:stochadeterflux} shows the time evolution of the open probability, $P_o(t)$, of the non-Markovian model and of the stochastic model for $\Lambda=2$ and $\tau_b=4$.
While both models converge to the same stationary state (with the same open probability and output fluxes), the relaxation is significantly different, except at short time (typically for $t<1/\Lambda$). In particular, for the non-stochastic model, the deterministic blockage release mechanism leads to slow, oscillatory relaxation towards the stationary state, whereas the relaxation is roughly exponential and faster for the stochastic model.


%
\end{document}